\documentclass{article} % default is 10 pt
\usepackage{subfigure}
\usepackage{graphicx} % needed for including graphics e.g. EPS, PS
\usepackage{pdfpages} % needed for including multi-page pdf documents
\usepackage{amssymb}
\usepackage{url}
\usepackage{appendix}
\usepackage{comment}
\usepackage{float}

\makeatletter
\def\@normalsize{\@setsize\normalsize{10pt}\xpt\@xpt
\abovedisplayskip 10pt plus2pt minus5pt\belowdisplayskip 
\abovedisplayskip \abovedisplayshortskip \z@ 
plus3pt\belowdisplayshortskip 6pt plus3pt 
minus3pt\let\@listi\@listI}

\def\subsize{\@setsize\subsize{12pt}\xipt\@xipt}
\def\section{\@startsection {section}{1}{\z@}{1.0ex plus
1ex minus .2ex}{.2ex plus .2ex}{\large\bf}}
\def\subsection{\@startsection 
   {subsection}{2}{\z@}{.2ex plus 1ex} {.2ex plus .2ex}{\subsize\bf}}
\makeatother
\begin{document}
\date{}

\title{\bf Constrained Simulation of the Bullet Cluster}

\author{Craig Lage and Glennys Farrar\\
Center For Cosmology and Particle Physics\\ 
Department of Physics\\
New York University, New York, NY 10003, USA\\
csl336@nyu.edu}

\maketitle

\subsection*{\centering Abstract}

% >>>>>>>>>>>>>>>>>>>>>>>>> Keywords and Abstract <<<<<<<<<<<<<<<<<<<<<
% Replace with your own keywords and abstract.  Text will be in italics
{\em Keywords: 
dark matter, simulation, astrophysics, galaxy clusters, X-ray clusters, gravitational lensing
}

In this work, we report on a detailed simulation of the Bullet Cluster (1E0657-56) merger, including magnetohydrodynamics, plasma cooling, and adaptive mesh refinement.  We constrain the simulation with data from gravitational lensing reconstructions and 0.5 - 2 keV Chandra X-ray flux map, then compare the resulting model to higher energy X-ray fluxes, the extracted plasma temperature map, Sunyaev-Zel'dovich effect measurements, and cluster halo radio emission.  We constrain the initial conditions by minimizing the chi-squared figure of merit between the full 2D observational data sets and the simulation, rather than comparing only a few features such as the location of subcluster centroids, as in previous studies.  A simple initial configuration of two triaxial clusters with NFW dark matter profiles and physically reasonable plasma profiles gives a good fit to the current observational morphology and X-ray emissions of the merging clusters.   There is no need for unconventional physics or extreme infall velocities.  The study gives insight into the astrophysical processes at play during a galaxy cluster merger, and constrains the strength and coherence length of the magnetic fields.  The techniques developed here to create realistic, stable, triaxial clusters, and to utilize the totality of the 2D image data, will be applicable to future simulation studies of other merging clusters.  This approach of constrained simulation, when applied to well-measured systems, should be a powerful complement to present tools for understanding X-ray clusters and their magnetic fields, and the processes governing their formation. 

\section{Introduction}
The mergers of clusters of galaxies are key events in the evolution of structure in our universe.  The ongoing Bullet Cluster merger (1E0657-56) is one of the most interesting such events known for several reasons, including its relatively simple structure and high surface brightness across the electromagnetic spectrum.  Because of this, the Bullet Cluster has been extensively studied observationally, making it an ideal laboratory for the study of the physics of galaxy clusters and their interactions.  The clear separation of the lensing mass centroids from the centroids of the X-ray emission has been taken as one of the strongest demonstrations of the reality of dark matter \cite{clowe06}, although this interpretation has been called into question and a modified gravity model has been proposed as an alternate explanation \cite{Moffat}.

We have undertaken to build a detailed simulation model of the Bullet Cluster merger, with the intent to use it to study cluster structure, dark matter-dark matter and dark matter-baryon interactions, possible modifications of general relativity, and to check the extent to which this cluster merger is or is not an outlier in $\rm \Lambda-CDM$ cosmology.  At the same time, constraints can be obtained on the present baryon and electron distributions, the magnetic field of the system and potentially on the non-thermal sources of support in X-ray clusters.

A number of simulation studies of the Bullet Cluster merger have been performed \cite{sf07}, \cite{Mastropietro}, \cite{milo2007}.  One of the improvements of this study compared to past studies is that, rather than compare the simulation to the data using a small number of extracted parameters (mass centroids, calculated velocities, etc.), we compare the simulation to the observational data on a pixel-by-pixel basis.  Extensive observations of the Bullet Cluster have been made at multiple wavelengths, and these two-dimensional images contain a large quantity of information.  Our approach makes use of this information to provide details of the structure of the initial clusters and the physics of the collision.  A second improvement presented here is the implementation of techniques to generate realistic, stable, triaxial clusters.  We will show that our techniques result in an excellent fit to the observed mass lensing distribution for this cluster collision.  We will also show reasonably good fits to the measurements of X-ray flux, S-Z effect, plasma temperature, and radio emission, although these results are more uncertain due to the additional ``gastrophysics'' degrees of freedom.

The paper is organized as follows.  We begin by reviewing the observational data that we use to constrain the simulations and the calculation of the figure of merit comparing the observations to the simulated images (Section \ref{Observations}).  This is followed by a description of the techniques for generating the triaxial clusters which are the inputs to the collision (Section \ref{Initial_Conditions}).  We then describe the optimization and error estimation techniques, show a set of image comparisons comparing the observations to the best fit simulations, and report what we have learned about the structure of the initial clusters (Section \ref{Results}).  Finally, we discuss some of the implications (Section \ref{Discussion_Section}), and conclude (Section \ref{Conclusions}). The Appendix describes the details of the simulations and describes how we calculate simulated images to compare to the observational data.

\section{Summary of Observations}
\label{Observations}
We compare six observational datasets to the simulation:
\vspace{-4 mm}
\begin{itemize} 
\renewcommand{\labelitemi}{$-$}
\itemsep0em
\item A mass lensing reconstruction
\item Three maps of X-ray intensity in different energy bins
\item The Sunyaev-Zel'dovich effect CMB temperature decrement
\item The radio halo intensity.  
\end{itemize}
\vspace{-4 mm}
Each dataset is converted to a 2D map containing 110x110 pixels, where for each pixel we have an observed value and an estimated uncertainty.  Two primary datasets, the mass lensing data and the lowest energy X-ray flux, are used to constrain the simulation initial conditions and fitting parameters.  The resulting simulation is then used to generate images which are compared to the remaining four datasets.

Using the two primary datasets, we construct the following combined figure of merit to measure the quality of fit between simulation and measurement  
	\begin{equation}
	\rm
	\chi^2 = \frac{1}{N_k N_i} \sum_{Observations = k}^{N_k}\hspace{0.2 in}\sum_{Pixels = i}^{N_i}\frac{(Sim^k_{i} - Obs^k_{i})^2}{(\sigma^k_{i})^2} ,
	\label{Chi2_calculation}
	\end{equation}

and then vary the parameters to minimize this $\rm \chi^2$.  For brevity, we refer to this parameter as $\chi^2$ throughout this work, but in fact it is $\rm \chi^2$ per degree of freedom.  We note that the parameter $\rm \chi^2$ is used as a figure of merit for finding the best fit initial conditions.  Our large-scale simulation provides a mean description of the system, and is incapable of capturing much of the small-scale physics such as inhomogeneous initial conditions and accretion of smaller mass distributions during the merging process.  For these reasons, although the $\rm \chi^2$ parameter is useful for evaluating the quality of the fit, we should not expect it to approach a value of one.

\subsection{Primary Constraining Datasets}

\begin{itemize}
	\item \begin{it}The gravitational lensing reconstruction from Bradac et.al. \cite{bradac06}.\end{it}  This dataset consists of the total projected mass in each 2D pixel as determined to reproduce the observed weak and strong lensing data.  The values of $\rm \sigma_i$ associated with the reconstruction have also been provided by M. Bradac [private communication]; these typically range from 5-25\% of the mass lensing data.

	\item \begin{it}X-ray flux measurements from the Chandra X-ray observatory \cite{Chandra}.\end{it}  A total of 9 separate observations are included in the datasets, representing a total observing time of 558 ks.  The X-ray flux is binned into three separate energy bins, 500eV - 2000eV, 2000eV - 5000eV, and 5000eV - 8000eV.  The lowest energy bin (500eV - 2000eV), which contains most of the photons, is our second primary dataset, along with the mass-lensing map.  The ``Ciao" \cite{Ciao} software analysis package is used to reduce the measured data to an X-ray flux in $\rm photons/(cm^2 sec)$.  For this data, a statistical uncertainty of $\rm 1/\sqrt{N_{photons}}$ and a systematic uncertainty of 1.7\% \cite{Chandra} are combined in quadrature to generate $\rm \sigma_i$.  
\end{itemize}

\begin {figure}[H]
	\centering
	\includegraphics[trim = 0.35in 2.5in 0.5in 0.90in, clip, width=\textwidth]{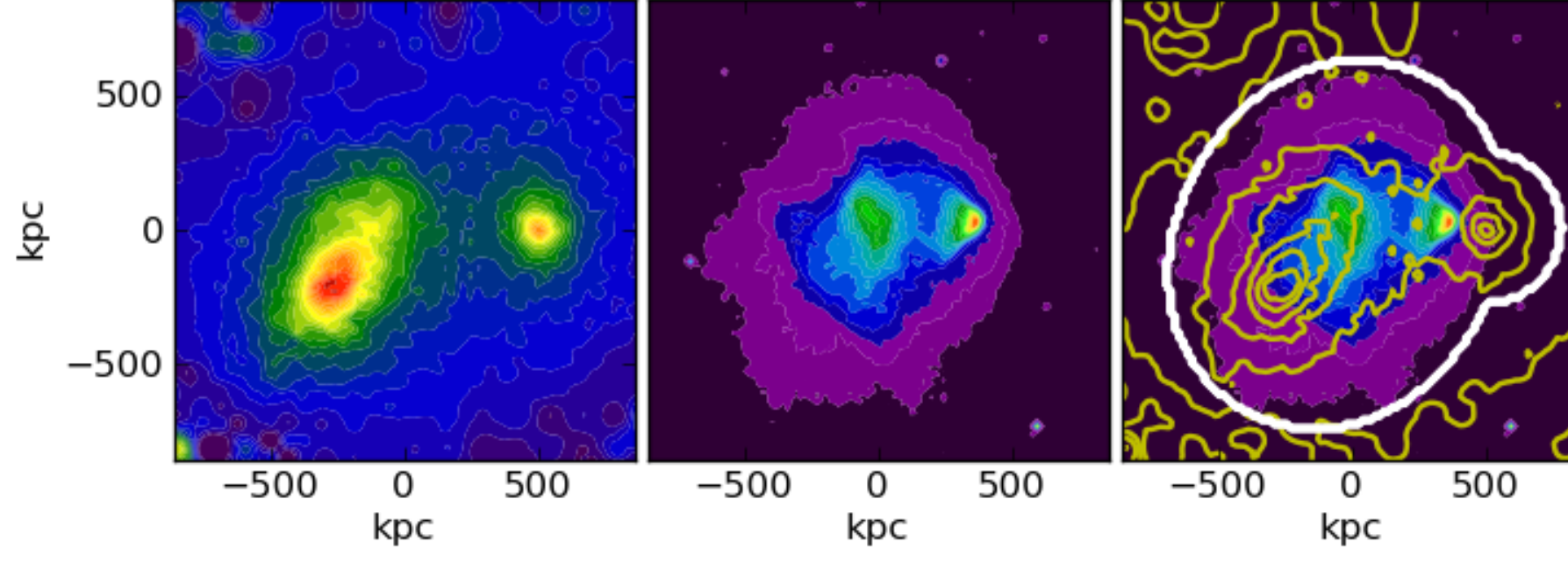}
	% Trim is Left Bottom Right Top
	\caption{Left: Mass lensing dataset.  Center: 500eV-2000eV X-ray flux dataset. Right: These two primary datasets overlaid. Only the region inside the heavy white outline is included in the $\chi^2$ calculation. }
	\label{Data_Overlay}
	% This figure is from bullet/data/kappa_23-Apr-12
  \end{figure}

We have manually identified the central region where the data has the highest confidence, and only pixels inside this region are included in the $\chi^2$ calculation in Equation \ref{Chi2_calculation}.  This region is shown by the heavy white outline in Figure \ref{Data_Overlay}; it contains approximately 5800 pixels in each dataset and is about 4.5 arc minutes across.

\subsection{Secondary Comparison Datasets}

\begin{itemize}
	\item \begin{it}The two higher energy X-ray bins (2000eV - 5000eV and 5000eV - 8000eV)\end{it} from the Chandra X-ray data, extracted as described above. 

%	\item \begin{it}Extracted plasma temperature [M. Markevitch - private communication, 2011].\end{it}  This is a map of the plasma temperature in each pixel, as extracted from the X-ray measurements and averaged along the line of sight.  Details of the extraction of this map are in \cite{Markevitch_TMap}.   The temperature $\rm \sigma_i$ [M. Markevitch - private communication] has a median value of 1.5 keV.

	\item \begin{it}A Sunyaev-Zel'dovich effect map from Plagge et.al. \cite{Plagge}.\end{it} This is a map of the SZE temperature decrement in $\rm \mu K$, measured using the South Pole Telescope. Based on \cite{Plagge}, $\rm \sigma_i$ is taken to have a constant value of 25 $\rm \mu K$. 

	\item \begin{it}Radio halo measurements at 1.3 GHz from Liang et.al. \cite{Liang}.\end{it} This dataset is a digitized version of the map in Figure 5 from \cite{Liang}. No attempt is made to assign a $\rm \sigma_i$ value for this dataset.
\end{itemize}

Plots of each of the six datasets, with corresponding simulation predictions, are shown in the following sections.

\section{Initial Conditions and Fitting Parameters}
\label{Initial_Conditions}
The simulations begin with two widely separated galaxy clusters, each in a state of dynamic equilibrium, approaching each other on a collision course.  For clarity, in what follows we will refer to the larger cluster as the main cluster, and the smaller cluster as the bullet cluster.  The bullet cluster initially approached the main cluster from the left, but has now passed through the main cluster and is currently on the right.  The major challenge in obtaining a simulation which produces a good fit to the observations lies in choosing appropriate initial conditions.

In the first stage of this effort we used spherically symmetric clusters.  This gave an approximate fit to the observations, but as our understanding increased we came to appreciate that some features of the system are most likely due to initial cluster triaxiality.  We find that initial clusters with a triaxial shape give a much better fit to the data, although this introduces additional variables.  With tens of thousands of measurements to fit, these additional parameters are in fact quite well constrained.  The procedures used to generate stable triaxial clusters with an NFW dark matter profile and a flexible baryon profile are described below.

We use a total of 34 fitting parameters, as listed in Table \ref{Fitting_Parameters} and described below, to model the collision.    

\subsection{Dark Matter Halos} 
For the dark matter halos, we assume that each of the initial clusters is described by a triaxial NFW profile\cite{NFW}, \cite{Lee_Suto}, with dark matter surfaces of constant density being a set of concentric ellipsoids, as follows:
	\begin{equation}
	\rm
	\rho_{DM} = \frac{\rho_{DM0}}{\frac{R}{R_C}(1+\frac{R}{R_C})^2} .
	\end{equation}
Here the radial parameter R is given by:
	\begin{equation}
	\rm
	R^2 = x^2 + \frac{y^2}{Q^2} + \frac{z^2}{P^2} ,
	\end{equation}
where P and Q are the triaxiality axis ratios.  We take $\rm P \le Q \le 1$, meaning that the x-axis is the long axis and the z axis is the short axis.  Each ellipsoid is then rotated to its initial orientation, as described later.  The parameters $\rm \rho_{DM0}$ and $\rm R_C$ can be written in terms of the virial radius $\rm R_{200}$, the concentration parameter C, the mass within the virial radius $\rm M_{200}$, and the critical density $\rm \rho_{CRIT}$ at redshift z as\cite{Lee_Suto}:
	\begin{equation}
	\rm
	R_C = \frac{R_{200}}{C} ,
	\end{equation}
	\begin{equation}
	\rm
	R_{200} = \left[\frac{M_{200} C^2}{4 \pi 200 \rho_{CRIT}(1+z)^3 (1+C)((1+C)\ln(1+C)-C)}\right]^{1/3} ,
	\end{equation}
	\begin{equation}
	\rm
	\rho_{DM0} = \frac{M_{200} C^3 (1+C)}{4 \pi R_{200}^3 ((1+C)\ln(1+C)-C)} .
	\end{equation}
There are thus a total of eight parameters to describe the two clusters: the mass $\rm M_{200}$, the concentration parameter C, and the shape parameters P and Q, for each of the two clusters.

\subsection{Baryonic Distributions} 
\label{Baryon_Profiles}
Following reference \cite{Lee_Suto}, we make the physically reasonable assumption that the density and temperature of the baryonic plasma are constant along surfaces of constant gravitational potential.  We find that correctly fitting the X-Ray emission data depends critically on the gas profiles of the initial clusters.  For this reason, we assume a flexible three-slope gas density profile, as follows:

	\begin{equation}
	\rm
	\rho_G = \frac{\rho_{G0}}{(1+(\frac{R}{R_{C1}})^2)^{\beta1}(1+(\frac{R}{R_{C2}})^2)^{\beta2-\beta1}(1+(\frac{R}{R_{C3}})^2)^{\beta3-\beta2}} ,
	\end{equation}

where the parameter R is given by $\rm R^2 = x^2 + y^2/Q(R)^2 + z^2/P(R)^2$, with P(R) and Q(R) the (slowly varying) shape parameters of the equipotential ellipsoids.  As described in detail in Reference \cite{binneyTremaine}, the equipotential ellipsoids defined by P(R) and Q(R) are aligned with the density ellipsoids, but they are more spherical than the density ellipsoids and become more spherical still as one moves out from the cluster center.

\begin{comment}
	\begin{equation}
	\rm
	R^2 = x^2 + \frac{y^2}{Q(R)^2} + \frac{z^2}{P(R)^2}
	\label{Gas_Equation}
	\end{equation}
\end{comment}

The central density parameter $\rm \rho_{G0}$ is adjusted so that the ratio of baryonic mass to total cluster mass within R200 is equal to an assumed gas fraction parameter GF.  This parameter GF is then taken as a fitting parameter for each cluster.  For given plasma and dark matter densities, the plasma temperature required for hydrostatic equilibrium is determined by evaluating the following integral for the particle internal energy
	\begin{equation}
	\rm
	u(R) = \frac{3}{2 \rho_G(R)} \int_R^{R_{max}}\frac{\partial \varphi}{\partial R'} \rho_G(R')dR'.
        \label{Temp_Calculation}
	\end{equation}

Since the plasma temperature is assumed constant along the equipotential ellipsoids, it is sufficient to evaluate this integral along the x-axis of the cluster, and use the resulting value all along the equipotential surface intersecting that axis at R.

There are thus 14 parameters needed to describe the baryonic matter distributions: the gas fraction parameter GF and three pairs of $\rm (R_C,\beta)$ parameters, for each of the two clusters.
\subsection{Cluster Generation Procedure}
  \label{Cluster_Generation_Procedure} 
\begin {figure}[H]
	\centering
	\subfigure[DM shape]
		  {\includegraphics[trim=0.9in 0.0in 1.0in 0.0in,clip,width=0.49\textwidth]{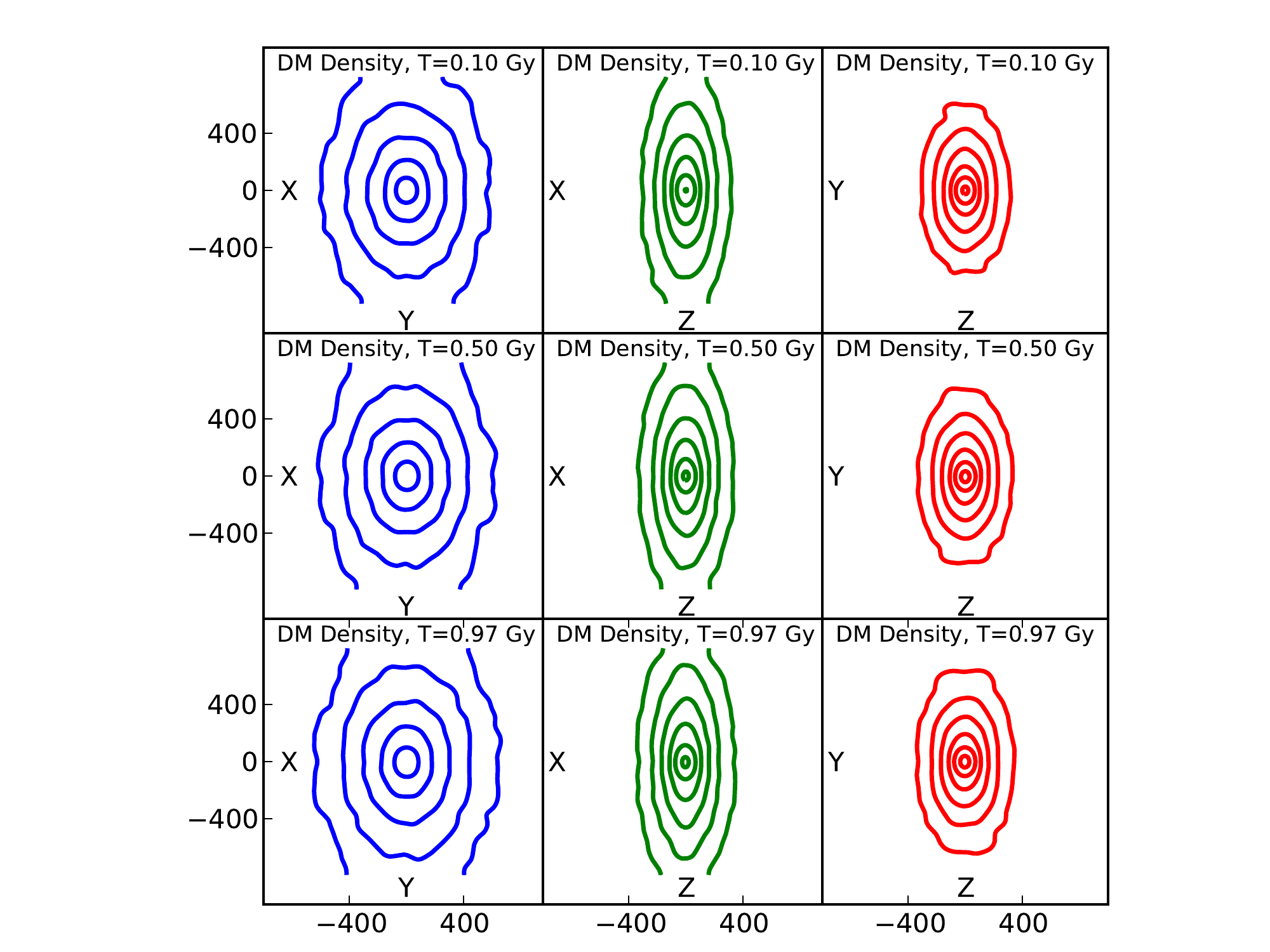}}
	\subfigure[Gas shape]
		  {\includegraphics[trim=0.9in 0.0in 1.0in 0.0in,clip,width=0.49\textwidth]{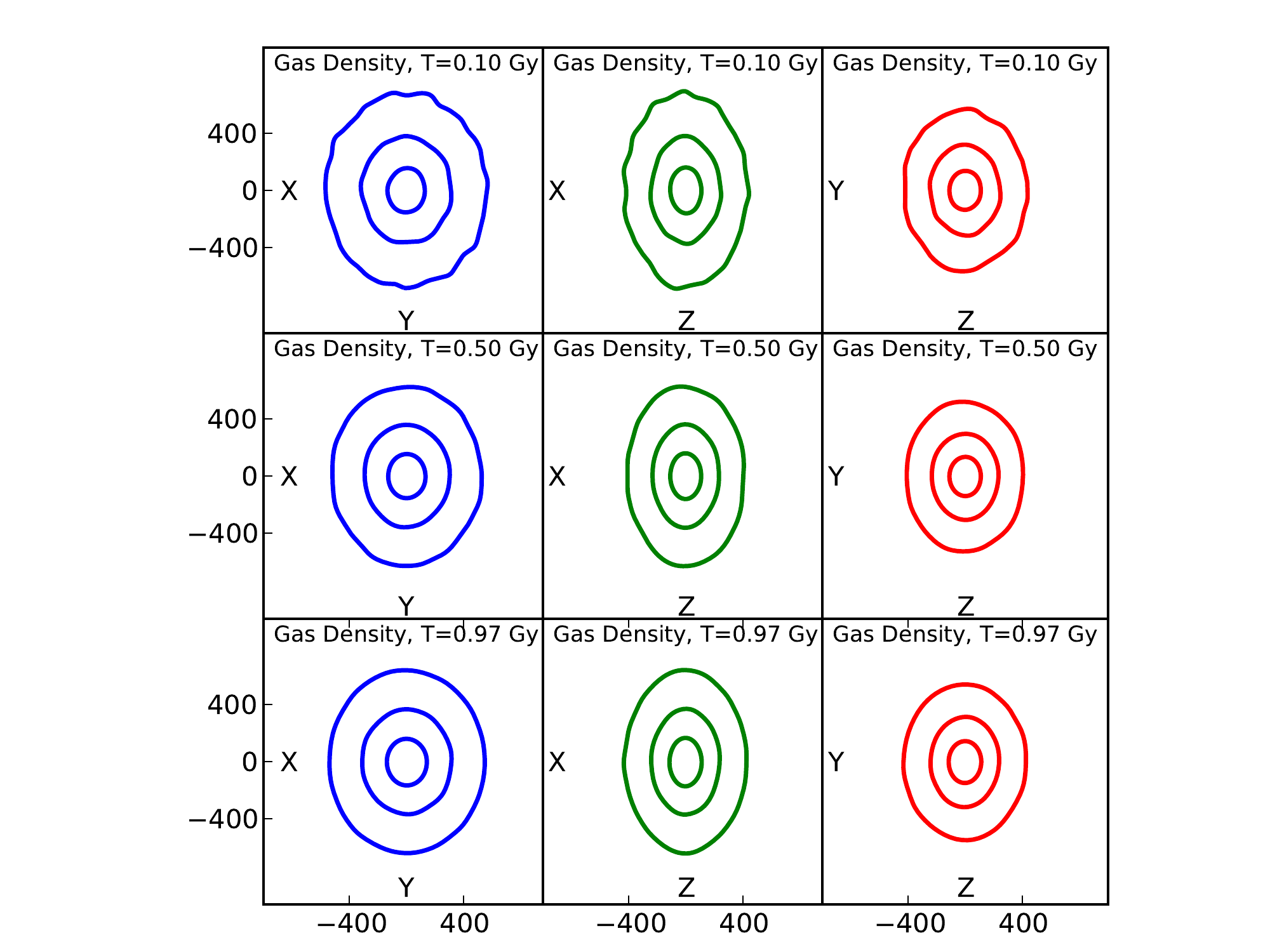}}
  \caption {Shape stability of a triaxial cluster with P = 0.35 and Q = 0.70}
  \label{Cluster_Stability_1}
  % Trim is Left Bottom Right Top
  % These graphs are from /bullet/code/enzo/profile_stability3/ddfiles/run3G.  They are created by the script in this directory.
  \end{figure}

\begin {figure}[H]
	\centering
	\includegraphics[trim=0.6in 0.6in 0.8in 0.4in,clip,width=0.8\textwidth]{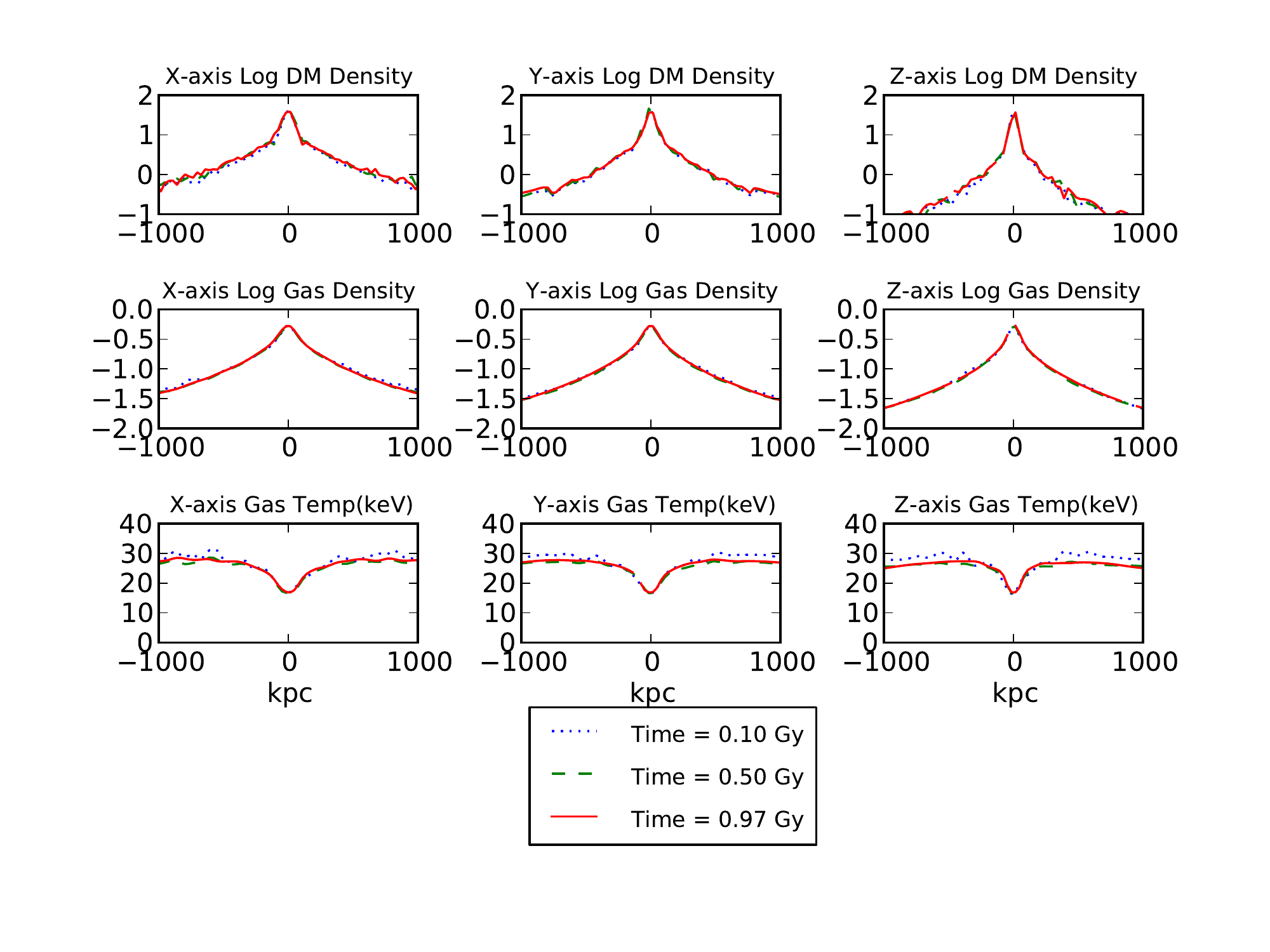}
  \caption {Profile stability of a triaxial cluster with P = 0.35 and Q = 0.70.  Note that the temperatures plotted are effective temperatures, as discussed in Sections \ref{Non-Thermal_Pressure_1} and \ref{Non-Thermal_Pressure_2}}
  \label{Cluster_Stability_2}
  % Trim is Left Bottom Right Top
  % These graphs are from /bullet/code/enzo/profile_stability3/ddfiles/run3G.  They are created by the script in that directory.
  \end{figure}

To combine these dark matter and gas profiles into a stable cluster we use the following procedure:
\begin{itemize}
	\item We choose the cluster mass, concentration parameter, and triaxiality parameters.  These parameters fix the mass density, and hence the gravitational potential of the cluster.
	\item We generate a stable dark matter halo using the Schwarzschild procedure\cite{Schwarzschild}.  This procedure involves assuming a randomly chosen initial position and velocity for each of a set of trial dark matter particles within the given potential, then calculating the orbit followed by each particle.  The density distribution which results from each particle following the calculated orbit is then determined.  A set of linear equations is solved to calculate the weight to be given to each of the trial particles in order to reproduce the original, assumed mass density distribution.  The dark matter particle initial positions and velocities are distributed along these orbits.  For these studies we use a total of 50,000 initial orbits, and several million dark matter particles (see Table \ref{Simulation_Conditions} in the Appendix).  We utilize a software package called SMILE \cite{Smile}, for carrying out the Schwarzschild procedure, and we find it to be very successful at generating stable triaxial halos.      
	\item We choose the parameters specifying the cluster baryon fraction (gas fraction) and the gas density profile. 
	\item Given the potential profile and the gas density profile, we calculate the gas temperature profile using Equation \ref{Temp_Calculation}.
        \item After each cluster is generated, we rotate it to an assumed orientation, specified by a total of six Euler angles, three for each cluster.
\end{itemize}
This procedure does involve an approximation, since the cluster gravitational potential is assumed set by the dark matter halo, whereas the cluster also contains a significant amount of baryonic matter.  In principle, we could iterate the procedure, calculating the potential due to the combined dark matter and baryonic mass distribution.  However, since the shape of the baryonic matter distribution is not too different from the shape of the dark matter distribution (see Figure \ref{Initial_Cluster_Profiles}), we find that it is sufficient to use the shape of the potential determined by the dark matter profile, but use the full cluster mass (dark matter + baryons) to set the magnitude of the potential.  Because the mass distribution profiles are similar, and the cluster mass is dominated by the dark matter, the error involved in this assumption is small and this procedure gives stable clusters.  We demonstrate the stability of clusters created by this procedure in Figures \ref{Cluster_Stability_1} and \ref{Cluster_Stability_2}, which show that a representative cluster is stable on a gigayear timescale.  Figure \ref{Cluster_Stability_1} shows the stability of the cluster shape, and Figure \ref{Cluster_Stability_2} shows the stability of the cluster profiles. 

\subsection{Initial Positions and Velocities} 
\begin {figure}[H]
	\begin{minipage}[b]{0.83 \textwidth}
	\centering
	\includegraphics[trim = 0.0in 2.0in 1.8in 2.0in, clip, width=\textwidth]{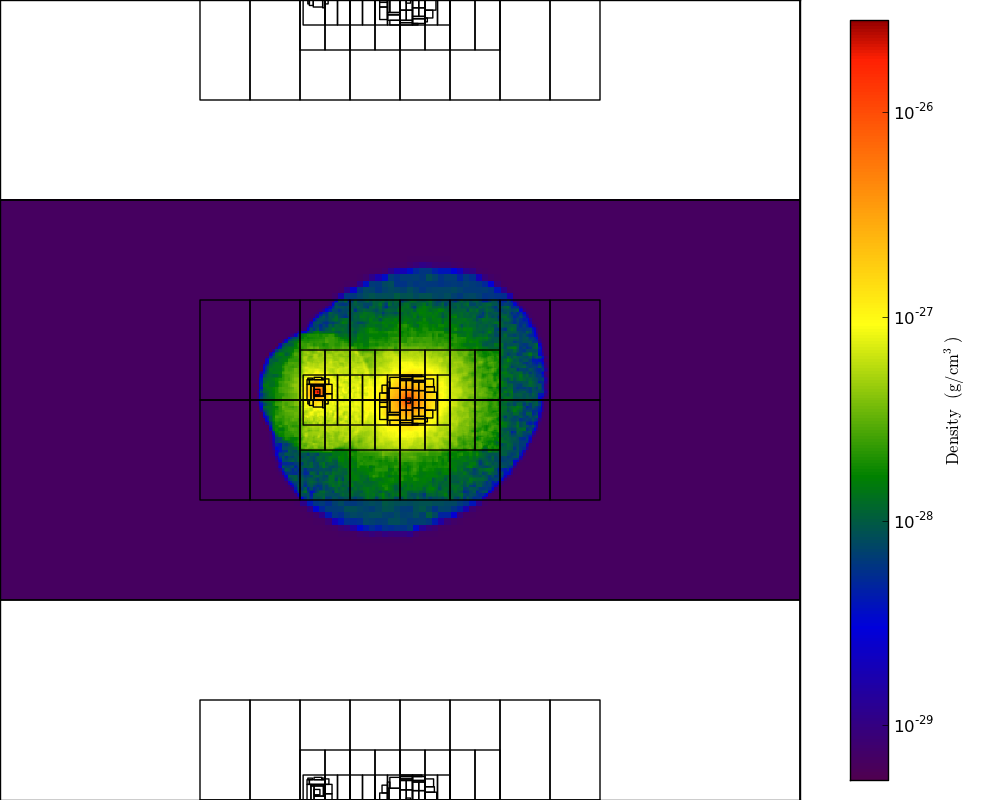}
	% trim is L B R T
	\end{minipage}
	\hspace{0.5cm}
	\begin{minipage}[b]{0.07 \textwidth}
	\centering
	\includegraphics[trim = 8.5in 0.0in 0.3in 0.0in, clip, width = \textwidth]{Initial_Slice.png}
	% trim is L B R T
	\end{minipage}
  \caption {This graph shows a typical baryon density slice near the beginning of the simulation. The black boxes show the initial mesh distribution. The size of the region shown is 12.0 x 6.0 Mpc.}
  \label{Initial_Density}
  % This graph is created by the program ytplots.py in the appropriate directory, and is called DD0000/output_0000_Slice_y_Density.png
  \end{figure}

The two initial triaxial clusters are assumed to fall in from infinity on a near-radial trajectory.  We begin the simulation when the virial radii of the two clusters make contact.  Figure \ref{Initial_Density} shows a density slice near the beginning of the simulation.  The initial velocities are controlled by two free parameters: the initial impact parameter of the two cluster mass centroids and a radial velocity percent increment over and above the velocity acquired while free-falling from infinite separation.  These are referred to as IP and Vinc respectively in Table \ref{Fitting_Parameters}.  We find the best fit value of Vinc to be about 10 percent less than unity, indicating that the clusters were initially bound, and also showing that large initial velocities are not required to reproduce the observations.

\subsection{Magnetic Field} 
\label{Mag_Field_1}
We set the initial magnetic field configuration of the two clusters as follows:
	\begin{itemize}
	\item We generate each of the three components of the Fourier-transformed B-fields $\rm \hat{B}_{x}, \hat{B}_{y}, \hat{B}_{z}$ as a Gaussian random field with a Kolmogorov spectrum ($\rm \hat{B_i} \propto k^{-5/3}$). The minimum and maximum possible k values for the initial random field configuration are given by: 
		\begin{equation}
		\rm
		k_{max} = \frac{2 \pi}{L} * \frac{N}{2}  \hspace{1 in} k_{min} = k_{max} / 4 ,
		\end{equation}
where L is the box length, which is 6 Mpc in these simulations, and N is the number of cells in the $x$-direction of the largest (coarsest) grid, which is 128.  Due to the adaptive mesh refined MHD, turbulence on smaller scales is generated as the simulation progresses.  An important physical question is what maximum coherence length characterizes the initial field configuration.  A few different choices have been explored in this first analysis, and we find that our results are sensitive to these choices.  The best results presented here use an initial maximum coherence length equal to four times the initial coarsest grid spacing, which is about 180 kpc.  Future work will explore this dependency in more detail.
	\item We clean the divergence in k-space by forcing $\rm \bf k\cdot \hat{B} = 0$.
	\item We then Fourier transform the B-field components back to real space.  All of the above steps are performed with the aid of the GarFields software package \cite{Garfield}.
	\item This generates a B-field of uniform magnitude throughout the simulation volume, whereas we expect the field to be stronger in regions of higher plasma density.  In a simple collapse model of a magnetized sphere, the density scales as $\rm1/r^3$ and the magnetic field scales as $\rm 1/r^2$.  We therefore scale the initial B-field magnitude so that $\rm |B| \propto \rho_{gas}^{2/3}$.  Note that the scaling factor is applied after the two clusters are combined into a single simulation file, so the same scaling factor (relationship between $\rm |B|$ and $ \rho_{gas}$) is used for both clusters.  This spatial scaling of the uniform Kolmogorov B-field introduces a slight non-zero value of $\rm \nabla \cdot B$.  We have verified that, because the length scale of the plasma density variation is much longer than the scale of B-field fluctuations, removing this $\rm \nabla \cdot B$ has negligible impact on the simulations, so we do not do this routinely.
	\end{itemize}

The only fitting parameter associated with the initial magnetic field configuration is thus an overall scale for the field magnitude, referred to as Mag in Table \ref{Fitting_Parameters}. This parameter is the peak magnetic field magnitude in the region of highest density in the initial configuration, which proves to be at the center of the bullet cluster.   

\subsection{Non-Thermal Pressure}
\label{Non-Thermal_Pressure_1}
In this simulation study we have attempted to calculate as many of the observables as possible from first principles, minimizing the number of ``fudge factors''.  However, we do find it necessary to include such an adjustable parameter to adequately describe the non-thermal pressure.  It has been known for some time (see, for example \cite{loeb94}) that a significant amount of non-thermal pressure support is needed in galaxy clusters in order to agree with the observations, and we find this as well.  We incorporate non-thermal pressure into this simulation using a single, space- and time-independent parameter, $\rm f_{ntp}$, following \cite{bode12}:
     \begin{equation}
       \rm
       P_{tot} = P_{th} + P_{ntp} = P_{th} (1 +\frac{f_{ntp}}{1-f_{ntp}}) = P_{th} \frac{1}{1-f_{ntp}} .
     \end{equation}

In effect, the simulation is performed with an effective temperature which is higher than the actual temperature in order to account for the increased pressure, then the effective temperature is reduced by the scaling factor when calculating X-ray flux and the associated plasma cooling.  This is a relatively primitive way of including non-thermal pressure and will be improved in future work.  Further discussion can be found in Section \ref{Non-Thermal_Pressure_2} where the predictions from the simulation are discussed.  

\subsection{Other Fitting Parameters} 
The remaining fitting parameters are:
\begin{itemize} 
\renewcommand{\labelitemi}{$-$}
\itemsep0em
\item The metallicity parameter Z (defined as a fraction of solar metallicity), which controls both the rate of cooling and the plasma X-ray emission, as discussed in Appendix \ref{Xray_Flux_Section}
\item The viscosity parameter, Visc, which adds a viscosity as a fraction of Spitzer viscosity, and will be discussed in more detail in Section \ref{Mag_Field_2}.
\end{itemize}

\section{Best Fit Initial Conditions}
\label{Results}
\subsection{Optimization}
\label{Optimization}
\begin{figure}[H]
	\centering
	\includegraphics[trim = 0.0in 1.0in 0.0in 1.8in, clip, width=\textwidth]{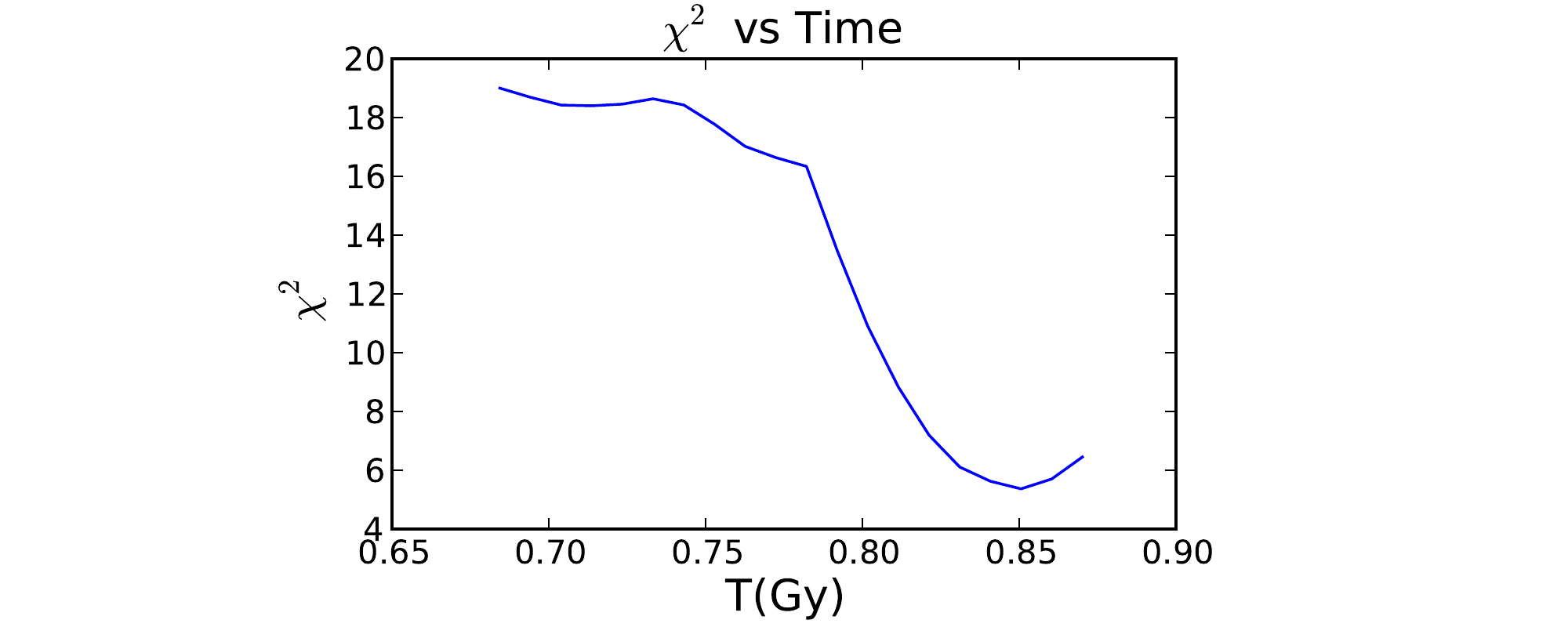}
	\caption{Typical evolution of the $\chi^2$ parameter (for the best fit viewing angle and position in the plane of the sky) as the simulation progresses.}
	\label{Chi_Squared_vs_Time}
	% This is from multi_matrixn76/ddfiles/run2D
	\end{figure}

For a given choice of parameters, the collision process is simulated as described in Appendix \ref{Simulation_Conditions_Section} and the full state of the system is recorded at time steps of 0.01 Gy in the relevant range of time.   The simulation is run in a frame with the initial cluster velocities in the x-direction, and the initial impact parameter in the y-direction, so there are two angular variables $\rm (\theta_{obs}, \psi_{obs})$ which determine how our viewing angle is related to the simulation coordinates.  For each time value T, a search is run through these viewing angles and the observables are calculated as described in Appendix \ref{Calculation_of_Observables}.  This generates a set of 2D images (one for each observable) for each set of values $\rm (T, \theta_{obs}, \psi_{obs})$.  These then need to be aligned to the observations in the plane of the sky, requiring three more variables $\rm (\Delta X, \Delta Y, \phi_{obs})$.  The values of $\rm (\Delta X, \Delta Y, \phi_{obs}, \theta_{obs}, \psi_{obs})$ which minimize the calculated $\chi^2$ value are determined for each value of T.  (After the initial conditions are approximately determined, this entire procedure only needs to be carried out for a limited time-range.)  Figure \ref{Chi_Squared_vs_Time} shows the typical evolution of the $\chi^2$ parameter through simulation time.  The observations are best described after approximately 0.85 Gy have elapsed since the beginning of the simulation.

We run simulations exploring the space of initial conditions and other parameters to find those which minimize the $\chi^2$ parameter.  These simulations are run on the NASA Pleiades supercomputer cluster, with each simulation running for about 8 hours of wall clock time.  About 10\% of this time is in setting up the initial conditions and analyzing the result.  The remainder is spent in running the Enzo simulation, typically using 64 CPUs.  The space of possible initial conditions is too large to carry out a fully systematized parameter optimization procedure, such as performing a Markov-Chain-Monte-Carlo.  Instead, Monte-Carlo searches are run within generously selected  ranges of parameters, with steepest-descent optimizations to find the locally-best fits for several of the best points in parameter space, supplemented by judicious by-hand exploration of parameter space to make sure no minima are overlooked.  No sampling procedure can absolutely guarantee one is near the global rather than just a local minimum for $\chi^{2}$, the strong constraints the data imposes make it likely that the best-fit initial conditions reported here are close to a global optimum, for the adopted model treatment.  In all, more than 1000 simulations were run in order to find the best fit initial conditions and their uncertainties given in Table \ref {Fitting_Parameters}.

There is some degree of decoupling in the parameter space.  The mass lensing projection is primarily determined by the dark matter distribution, and hence is mostly determined by the shapes and orientations of the dark matter halos, which are controlled by the parameters labeled ``Dark Matter Halo Parameters'' and ``Orbital Geometry Parameters'' in Table \ref{Fitting_Parameters}.  To most efficiently find the global minimum, the optimization strategy we follow is first to optimize these parameters with a $\chi^2$ calculated only from the mass lensing data, then optimize the remaining parameters with a $\chi^2$ calculated from both the mass lensing data and the lowest energy X-ray flux, and finally to optimize on the full parameter set with a $\chi^2$ calculated from both the mass lensing data and the X-ray flux data.

\subsection{Estimation of Parameter Uncertainties}
\label{Statistics_Section}
To estimate the uncertainty associated with the parameters determined from the optimization, one would ideally perform a Markov Chain Monte Carlo (MCMC) analysis of the simulation model in the multi-dimensional space of initial conditions.  However, the minimal such MCMC analysis for a system such as ours involves running tens or hundreds of thousands of trials, and the simulation is too computationally expensive to allow this.  The strategy we use is to run a smaller number of trials, build an approximate model of $\chi^2$ in the multi-dimensional parameter space, then characterize the parameter distributions using this model.  This procedure is described in more detail in Blizniouk, et.al. \cite{Blizniouk}, where it is shown to give distributions similar to that which result from running an MCMC analysis on the original computationally expensive simulation.

Simulations are run using a range of initial conditions, and a $\chi^2$ value is calculated for each.  Some of these simulation runs (approximately 700) are part of the $\chi^2$-minimization runs, and some (approximately 300) are run with intentionally varied parameters in order to span the space of input parameters.  Many of these simulations are run with the lower resolution conditions described in the Appendix.  We then use the results of these simulation runs to build a multi-dimensional cubic-spline Radial Basis Function (RBF) model of $\chi^2$ as a function of the input parameters.  This RBF model fits the simulated points exactly, and varies smoothly as one moves away from the simulated points.  

\begin{figure}[H]
	\centering
	\includegraphics[trim=0.1in 2.5in 0.1in 0.6in,clip,width=\textwidth]{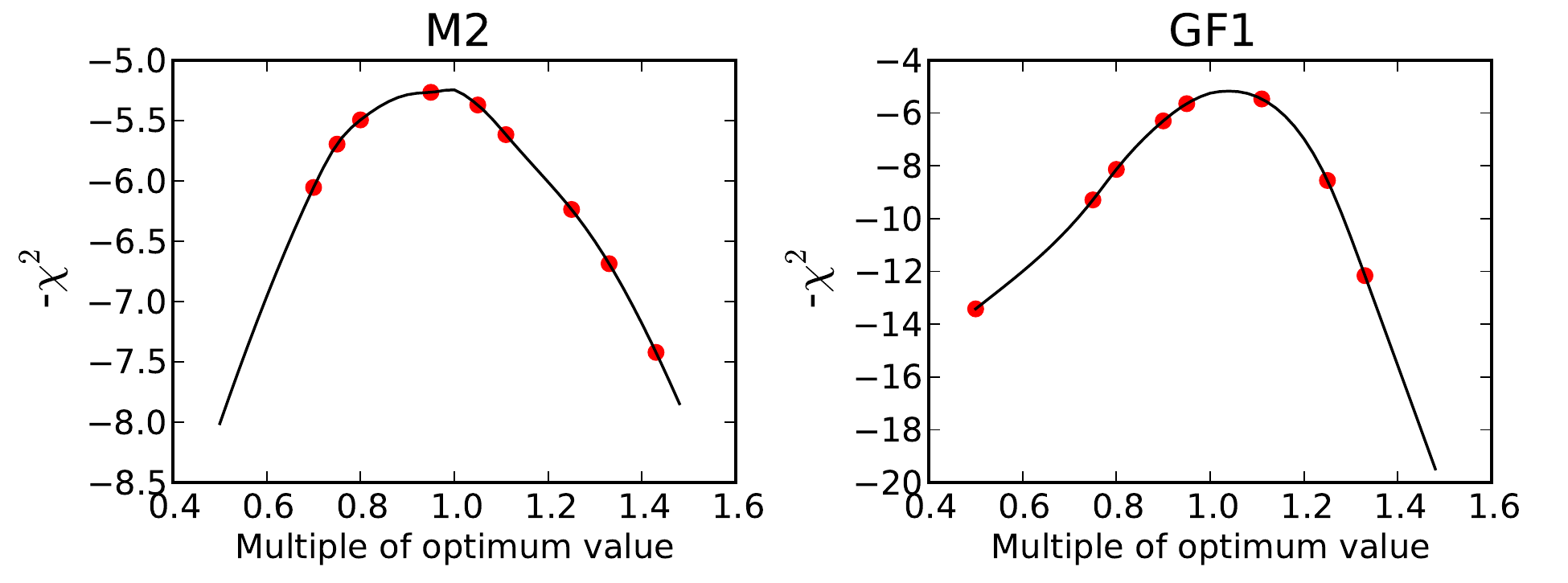}
	% Trim is Left Bottom Right Top
	\caption{Behavior of the Radial Basis Function model as two representative parameters M2(Bullet Cluster mass) and GF1(Main Cluster gas fraction) are varied around the optimum point, with all other parameters held fixed.  The RBF model is the black lines, and the actual simulations are the red circles.}
	\label{Model_Fit}
	\end{figure}

Figure \ref{Model_Fit} shows plots of the RBF model as two typical parameters move away from the optimum point with all other parameters held fixed.  The RBF model, which is computationally easy to evaluate, is then used to estimate the uncertainties of the parameters, defined so that the region within 1-sigma on either side of the best-fit value of a given parameter contains 34\% of the probability density after marginalizing over all other parameters.  The sigma values which result are tabulated in Table \ref{Fitting_Parameters}. 

% Statistical Model Fit
% These statistics runs are in full_multi_lowresn132 and multi_lowresn140.  Approx 370 runs total with each of 37 parameters multiplied by
% (0.5X, 0.75X, 0.8X, 0.9X, 0.95X, 1.05X, 1.1X, 1.2X, 1.5X, 2.0X)
% Analysis is in pleiades_stuff/paper_stats

\begin{table}[H]   
	\centering
	\begin{tabular}{|c|c|c|c|c|} 
		\hline 
		\multicolumn{5}{|c|}{Fitting Parameters with Best Fit Values} \\ 
		\hline 
		\multicolumn{5}{|c|}{Dark Matter Halo Parameters} \\ 
		\hline 
		Parameter & Description & Value & Sigma & Units \\ 
		\hline 
		$\rm M_1$ & Main Cluster Mass ($\rm M_{200}$) & 1.91E15 & 0.20E15 & $\rm M_\odot$ \\ 
		\hline 
		$\rm M_2$ & Bullet Cluster Mass ($\rm M_{200}$) & 2.59E14 & 0.31E14 &  $\rm M_\odot$ \\ 
		\hline 
		$\rm C_1$ & Main Cluster Concentration & 1.17 & 0.14 &  - \\ 
		\hline 
		$\rm C_2$ & Bullet Cluster Concentration & 5.45 & 0.70 &  - \\ 
		\hline
		$\rm P_1$ & Main Cluster Z/X Axis Ratio & 0.35 & 0.05 &  - \\ 
		\hline 
		$\rm Q_1$ & Main Cluster Y/X Axis Ratio & 0.68 & 0.09 &  - \\ 
		\hline
		$\rm P_2$ & Bullet Cluster Z/X Axis Ratio & 0.61 & 0.08 &  - \\ 
		\hline 
		$\rm Q_2$ & Bullet Cluster Y/X Axis Ratio & 0.68 & 0.10 &  - \\ 
		\hline 
		\multicolumn{5}{|c|}{Gas Profile Parameters} \\ 
		\hline 
		Parameter & Description & Value & Sigma & Units \\ 
		\hline 
		$\rm GF_1$ & Main Cluster Gas Fraction & 0.19 & 0.02 &  - \\ 
		\hline 
		$\rm GF_2$ & Bullet Cluster Gas Fraction & 0.17 & 0.02 &  - \\ 
		\hline
		$\rm RC_{11}$ & Main Cluster Gas Radius1 & 59.4 & 7.9 &  kpc \\ 
		\hline 
		$\rm RC_{12}$ & Bullet Cluster Gas Radius1 & 19.8 & 1.9 &  kpc \\ 
		\hline
		$\rm \beta_{11}$ & Main Cluster Exponent1 & 0.38 & 0.06 &  - \\ 
		\hline 
		$\rm \beta_{12}$ & Bullet Cluster Exponent1 & 0.51 & 0.07 &  - \\ 
		\hline
		$\rm RC_{21}$ & Main Cluster Gas Radius2 & 69.9 & 11.4 &  kpc \\ 
		\hline 
		$\rm RC_{22}$ & Bullet Cluster Gas Radius2 & 47.8 & 6.4 &  kpc \\ 
		\hline
		$\rm \beta_{21}$ & Main Cluster Exponent2 & 0.45 & 0.05 &  - \\ 
		\hline 
		$\rm \beta_{22}$ & Bullet Cluster Exponent2 & 0.85 & 0.14 &  - \\ 
		\hline
		$\rm RC_{31}$ & Main Cluster Gas Radius3 & 647 & 82 &  kpc \\ 
		\hline 
		$\rm RC_{32}$ & Bullet Cluster Gas Radius3 & 465 & 80 &  kpc \\ 
		\hline
		$\rm \beta_{31}$ & Main Cluster Exponent3 & 0.67 & 0.05 &  - \\ 
		\hline 
		$\rm \beta_{32}$ & Bullet Cluster Exponent3 & 0.50 & 0.06 &  - \\ 
		\hline
		\multicolumn{5}{|c|}{Orbital Geometry Parameters} \\ 
		\hline 
		Parameter & Description & Value & Sigma & Units \\ 
		\hline 
		$\rm \phi_1$ & Main Cluster Euler Angle 1 & 185 & 33 &  Degrees \\ 
		\hline 
		$\rm \theta_1$ & Main Cluster Euler Angle 2 & 38.4 & 5.9 &  Degrees \\ 
		\hline 
		$\rm \psi_1$ & Main Cluster Euler Angle 3 & 221 & 30 &  Degrees \\ 
		\hline 
		$\rm \phi_2$ & Bullet Cluster Euler Angle 1 & 164 & 23 &  Degrees \\ 
		\hline 
		$\rm \theta_2$ & Bullet Cluster Euler Angle 2 & 100 & 14 &  Degrees \\ 
		\hline 
		$\rm \psi_2$ & Bullet Cluster Euler Angle 3 & 65.0 & 10 &  Degrees \\ 
		\hline 
		$\rm IP$ & Impact Parameter & 256 & 35 &  kpc \\ 
		\hline
		$\rm V_{Inc}$ & Infall Velocity Increment & -10.9 & 15 &  \% \\ 
		\hline
		\multicolumn{5}{|c|}{Remaining Parameters} \\ 
		\hline 
		Parameter & Description & Value & Sigma & Units \\ 
		\hline 
		$\rm Z$ & Metallicity (Cooling) & 0.78 & 0.10 &  Solar \\ 
		\hline
		Mag & Peak Magnetic Field Magnitude & 61.0 & 5.4 &  $\rm \mu G$ \\ 
		\hline
		$\rm f_{ntp}$ & Non-Thermal Pressure factor & 0.52  & 0.09 & - \\ 
		\hline
		$\rm Visc$ & Viscosity & 0.12  & 0.02 & Fraction of Spitzer \\ 
		\hline
	\end{tabular} 
	\caption{Best fit parameters determined from the simulations, as well as an estimate of the uncertainties.  The determination of the uncertainties is described in Section \ref{Statistics_Section}}
	% These parameters are from multi_matrixn76/run2
	\label{Fitting_Parameters}
\end{table}

\subsection{Comparison of Best Fit Simulation to Observations}
\label{Comparison}
Following the procedures in the preceding sections leads to the best fit initial conditions summarized in Table \ref{Fitting_Parameters}.  This section discusses a series of images which exemplify the fit between the optimized simulation and the observations.   We begin with results which are governed mainly by the initial conditions on the dark matter and the constraints from mass-lensing data.  Then we turn to the more ``gastrophysics''-dependent aspects.  

\subsubsection{Mass Lensing}
\label{MassLensing}
\begin{figure}[H]
	\centering
	\subfigure[Fit achieved with $\chi^2$ calculated from mass lensing data only.  Mass only $\chi^2 = 1.15$.]{\includegraphics[trim = 1.0in 0.0in 0.7in 0.34in, clip, width=0.48\textwidth]{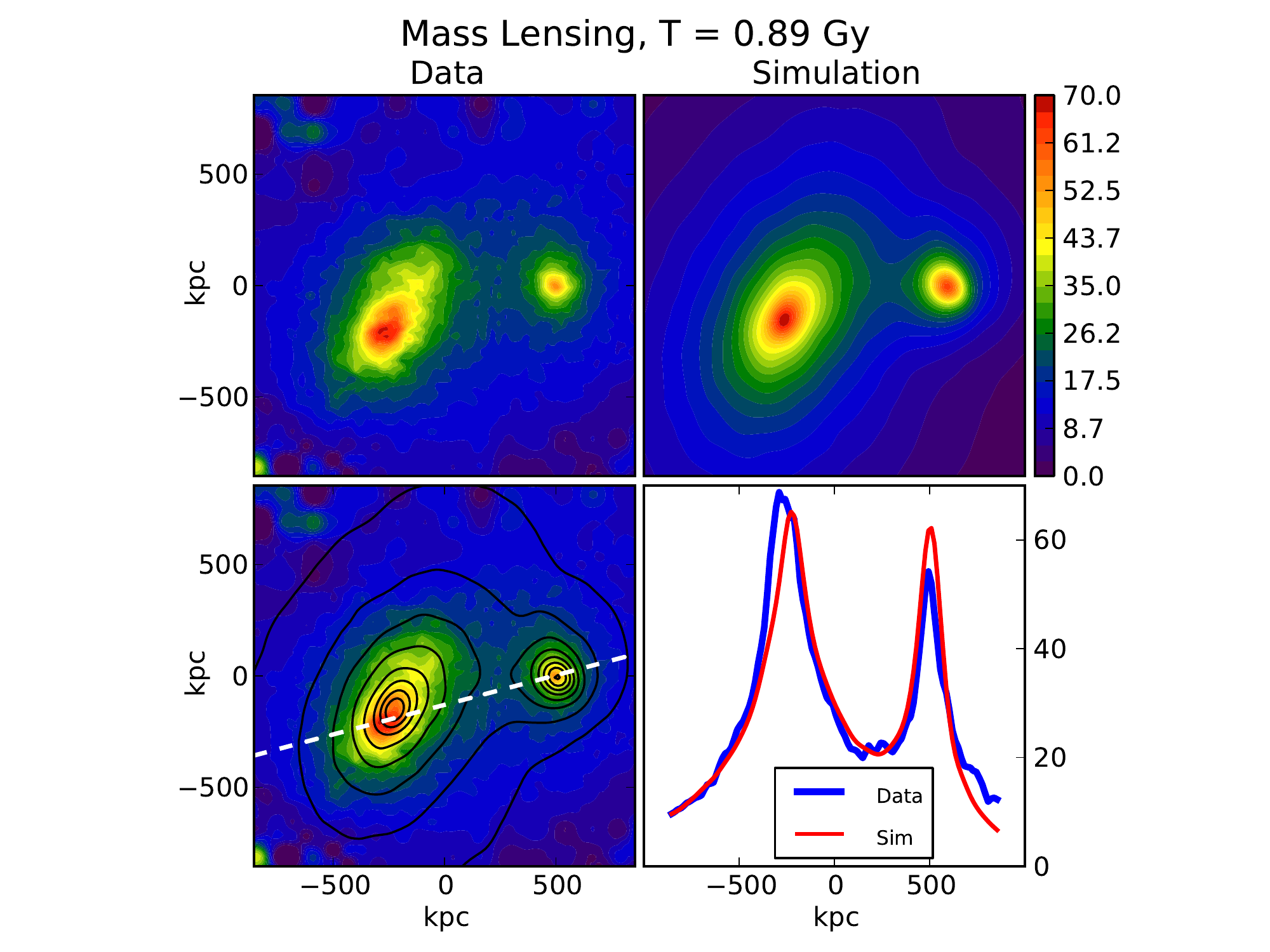}\label{Mass_only}}
	\subfigure[Fit achieved with $\chi^2$ calculated from mass lensing data and lowest-energy X-ray flux data.  Mass lensing contribution to $\chi^2 = 2.04$.]{\includegraphics[trim = 1.0in 0.0in 0.7in 0.34in, clip, width=0.48\textwidth]{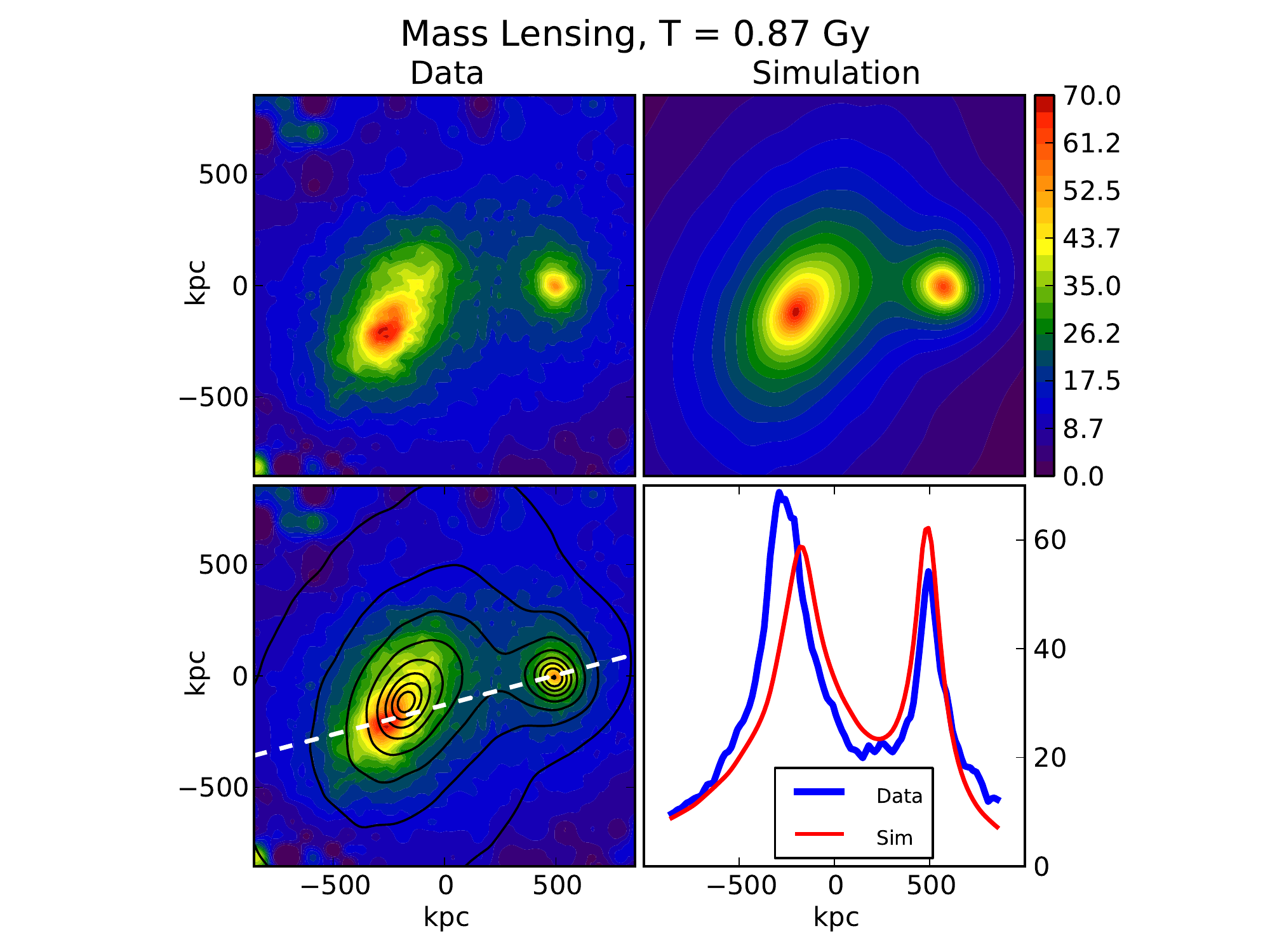}\label{Best_Fit_Lensing}}
	\vspace{-0.15in}
	\caption{Mass lensing fit between the data and the simulation.  In each of these plots, the measured data is in the upper left and the simulated result, on the same scale, is in the upper right.  The lower left shows an overlay of the measured data and the simulation, with the simulation shown as black contour lines, and the lower right shows the measured data and simulation along a line through the 2D data planes chosen to pass approximately through the measured peaks; this slice is shown as a dotted white line in the lower left.}
	% This plot is from bsearch_matrixn110/batchfiles/batch3/ddfiles/run4 optimizing on mass only
	% trim is L B R T
  \end{figure}

Figure \ref{Mass_only} shows the predicted mass-lensing map when parameters are optimized using $\chi^2$ calculated only from the mass lensing data, showing that the model is quite successful at reproducing the mass lensing distributions. The initial cluster triaxiality reproduces the shapes of the clusters quite well, and the value of $\chi^2 = 1.15$ obtained shows a good fit to the observations.  For comparison, Figure \ref{Best_Fit_Lensing} shows the fit of the same simulation with a $\chi^2$ calculated from both the mass lensing data and the lowest energy X-ray flux.  Since Figure \ref{Mass_only} and  Figure \ref{Best_Fit_Lensing} are from the same simulation (using the parameters in Table \ref{Fitting_Parameters}), only the alignment to the observational data is different between these two figures.  The value of $\chi^2 = 2.04$ obtained in Figure \ref{Best_Fit_Lensing} shows that the quality of the fit using only mass lensing data is degraded slightly when the alignment is chosen to give the best simultaneous fit including the X-ray flux as described below. 

\begin {figure}[H]
    \subfigure[]{\includegraphics[trim=3.0in 2.0in 3.0in 2.0in,clip,width=0.48\textwidth]{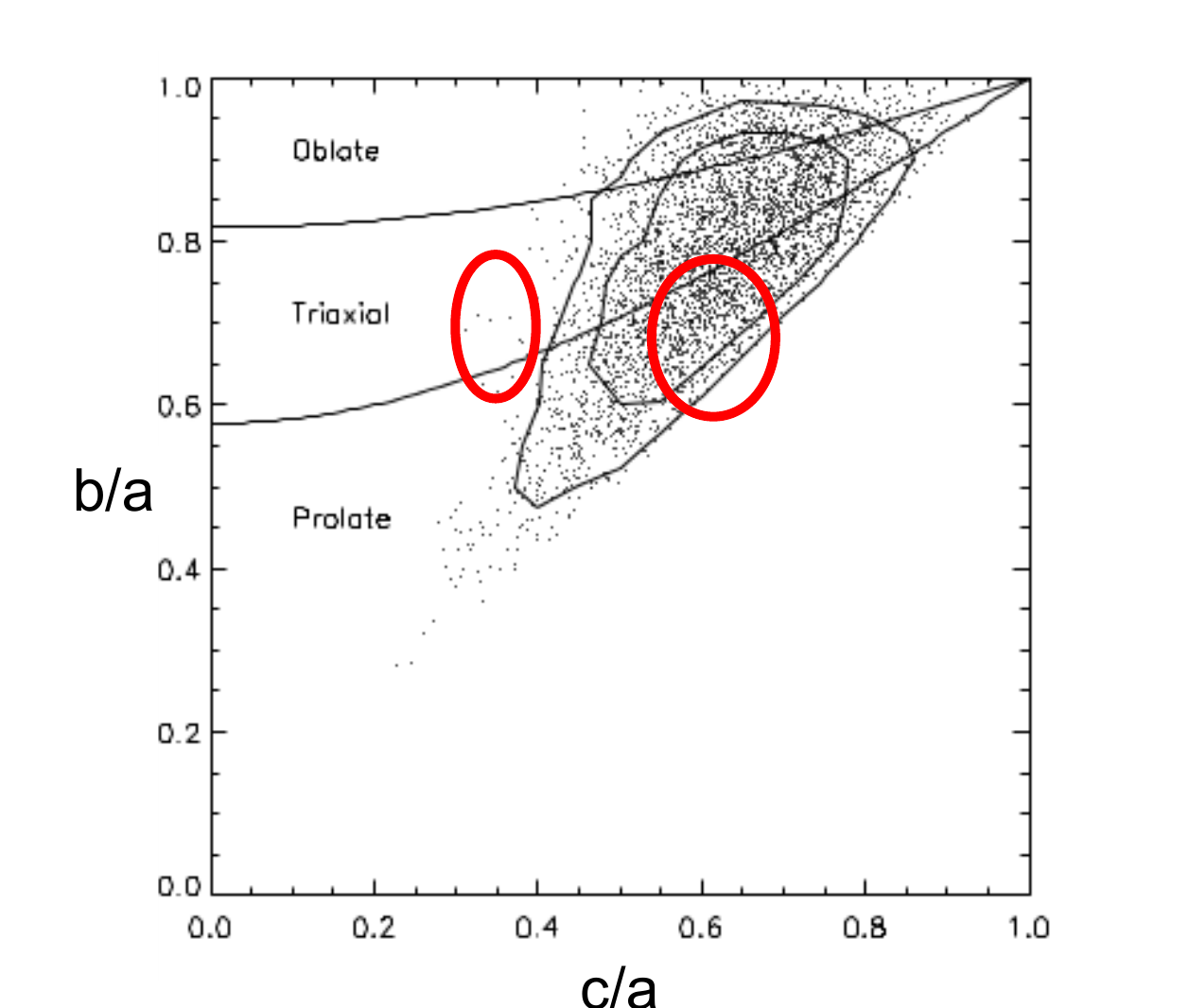}\label{Triaxialities}}
    \subfigure[]{\includegraphics[trim = 2.0in -1.0in 4.0in 1.0in, clip, width=0.48\textwidth]{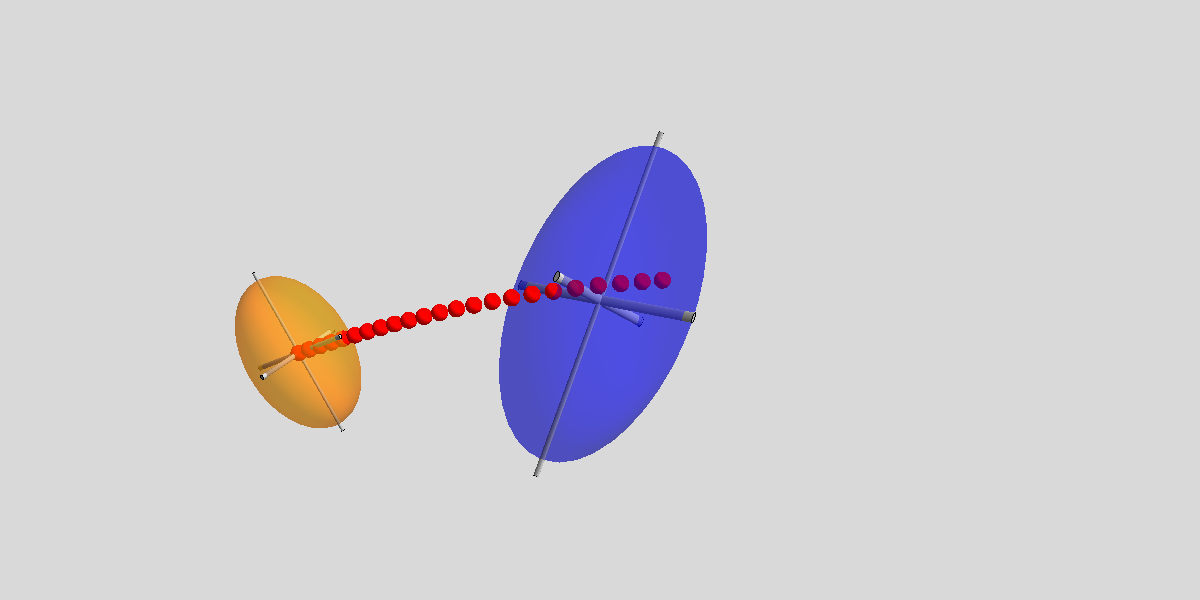}\label{Collision_Path}}
    	\vspace{-0.15in}
    \caption {Figure (a), reproduced from Bailin and Steinmetz \cite{bailin2005}, shows the best-fit triaxiality ratios of the two clusters, as compared to N-body simulations (dominated by lower mass clusters), with our simulation results overlaid as red ellipses; $\rm c/a=P~ and ~ b/c = Q$ in the notation of Table  \ref{Fitting_Parameters}.  The extent of the ellipses are 1-$\sigma$ errors; the left-hand ellipse is the main cluster, while the right-hand ellipse is the bullet cluster.  Figure (b) shows the path of the collision, as seen in the plane of the sky, illustrating the initial orientations of the two ellipsoids.  The red dots indicate the path of the dark matter centroids.}
  \end{figure}

Figure \ref{Triaxialities}, reproduced from Bailin et.al. \cite{bailin2005}, shows how the best fit triaxiality parameters given in Table \ref{Fitting_Parameters} compare to those found in an analysis of large N-body simulations of structure growth.  While the bullet cluster is well within the normal population, the large triaxiality range of the main cluster appears to make it much more unusual.  This large asymmetry may reflect a prior merger which took place in this cluster.  However, Schneider et.al. \cite{Schneider2012} have reported that the ellipticity of galaxy clusters increases as the mass increases, up to a mass of $\rm 2\times 10^{14} M_{\odot}$, so that our value of $P \equiv c/a = 0.35$ for the massive main cluster seems not to be an outlier.  This and other aspects of the Dark Matter initial conditions will be discussed in the context of $\Lambda$CDM cosmology in a companion paper, in preparation. 

Figure \ref{Collision_Path} shows the collision as viewed from our perspective, and is intended to help visualize the orientations of the cluster ellipsoids.  The best-fit relative velocity vector between the bullet cluster and the main cluster is inclined approximately 10 degrees to the plane of the sky.  This best fit radial velocity has the bullet cluster dark matter centroid receding from us at 837 km/sec relative to the main cluster dark matter centroid.  This is to be compared with the radial velocity analysis of Barrena, et.al. \cite{barrena}, who found that the bullet subcluster galaxies have a velocity offset of 616 $\pm$ 80 km/sec relative to the main cluster galaxies in the main cluster's rest frame.  Our simulation makes a prediction for the distribution of velocities and their variances as a function of position in the sky, so a more detailed comparison to the current full dataset is warranted to determine whether the discrepancy (about $\rm 2.7 \sigma$ with the Barrena et.al. errors) is significant.  

\subsubsection{X-ray Flux}
\begin{figure}[H]
	\centering
	\subfigure[500-2000eV]{\includegraphics[trim = 1.0in 0.0in 1.5in 0.34in, clip, width=0.331\textwidth]{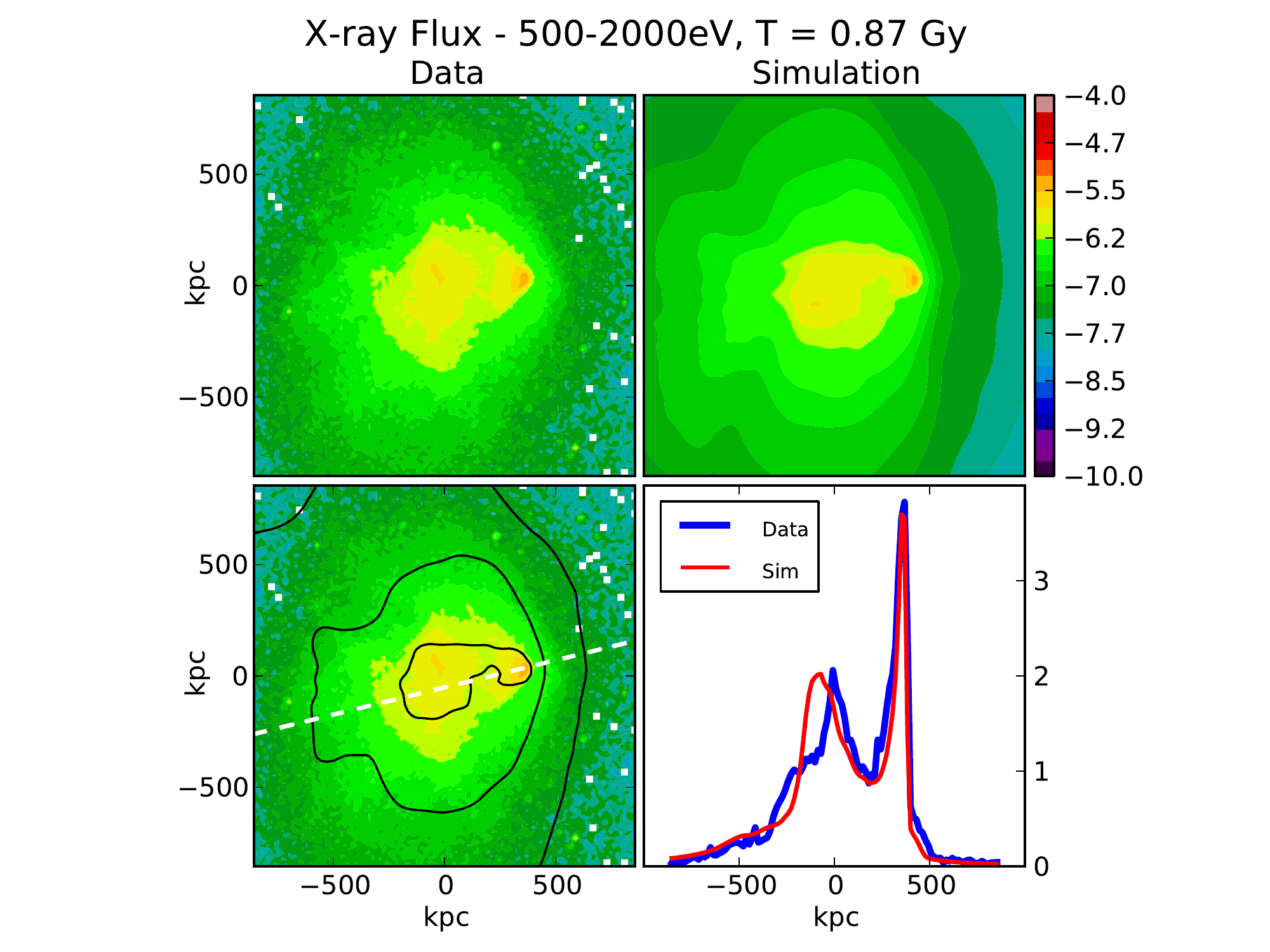}\label{Best_Fit_X-ray1}}
	\subfigure[2000-5000eV]{\includegraphics[trim = 1.58in 0.0in 1.5in 0.34in, clip, width=0.296\textwidth]{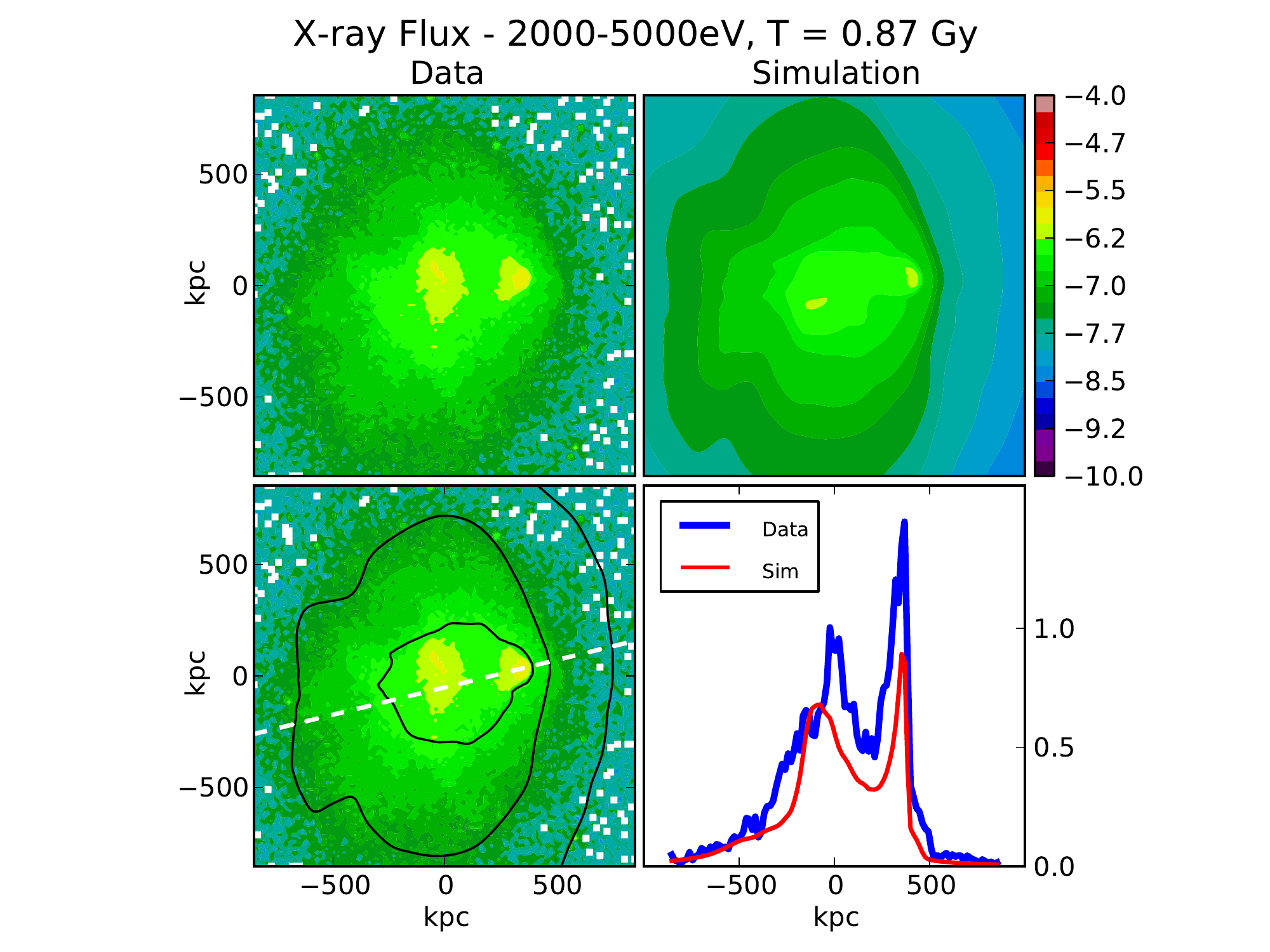}\label{Best_Fit_X-ray2}}
	\subfigure[5000-8000eV]{\includegraphics[trim = 1.58in 0.0in 0.7in 0.34in, clip, width=0.344\textwidth]{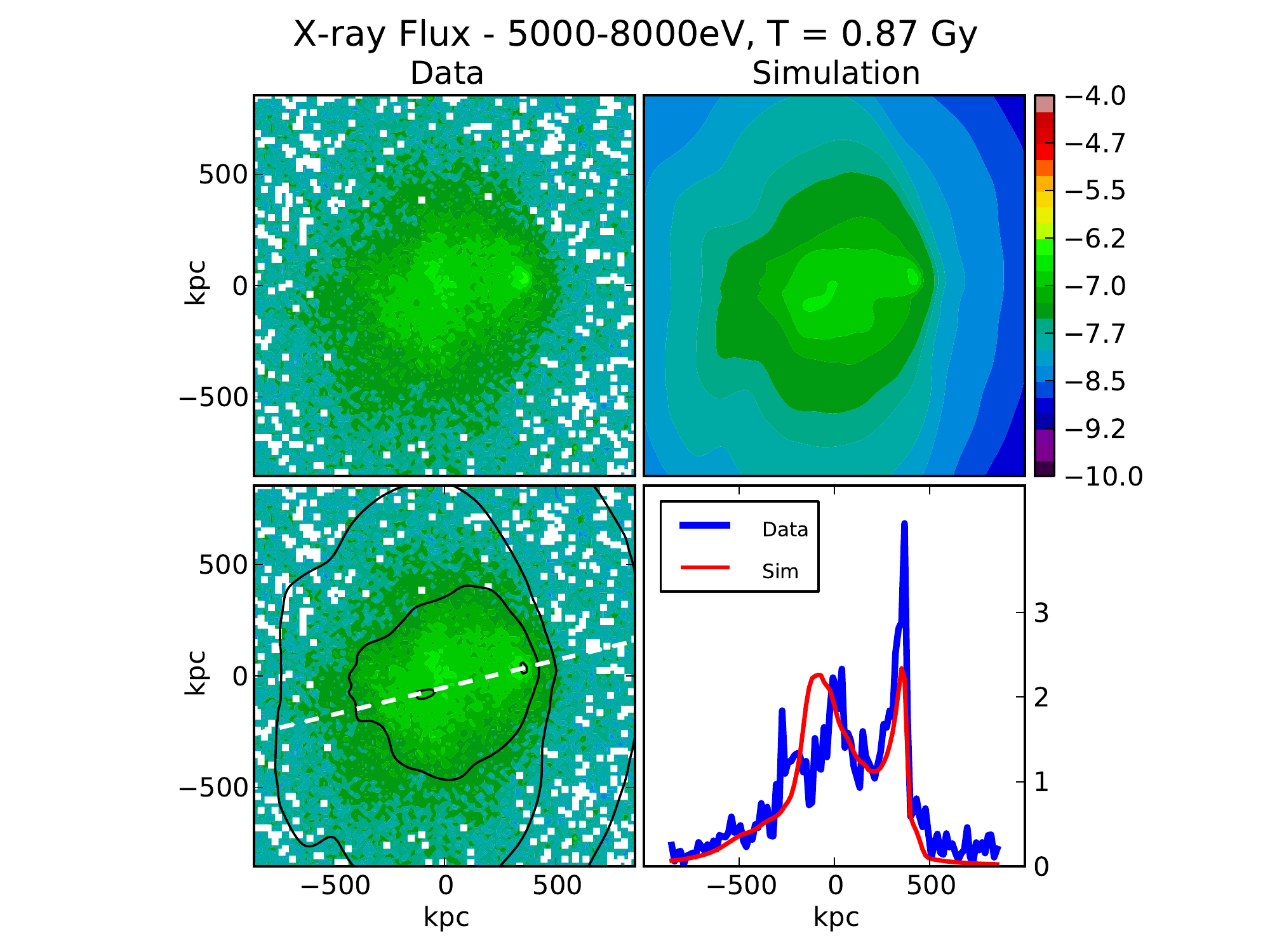}\label{Best_Fit_X-ray3}}
	\caption{Best fit result for the X-ray flux.}
\end{figure}

\begin{figure}[H]
	\centering
	\subfigure[500-2000eV Log scale]{\includegraphics[trim = 1.0in 0.0in 0.7in 0.34in, clip, width=0.48\textwidth]{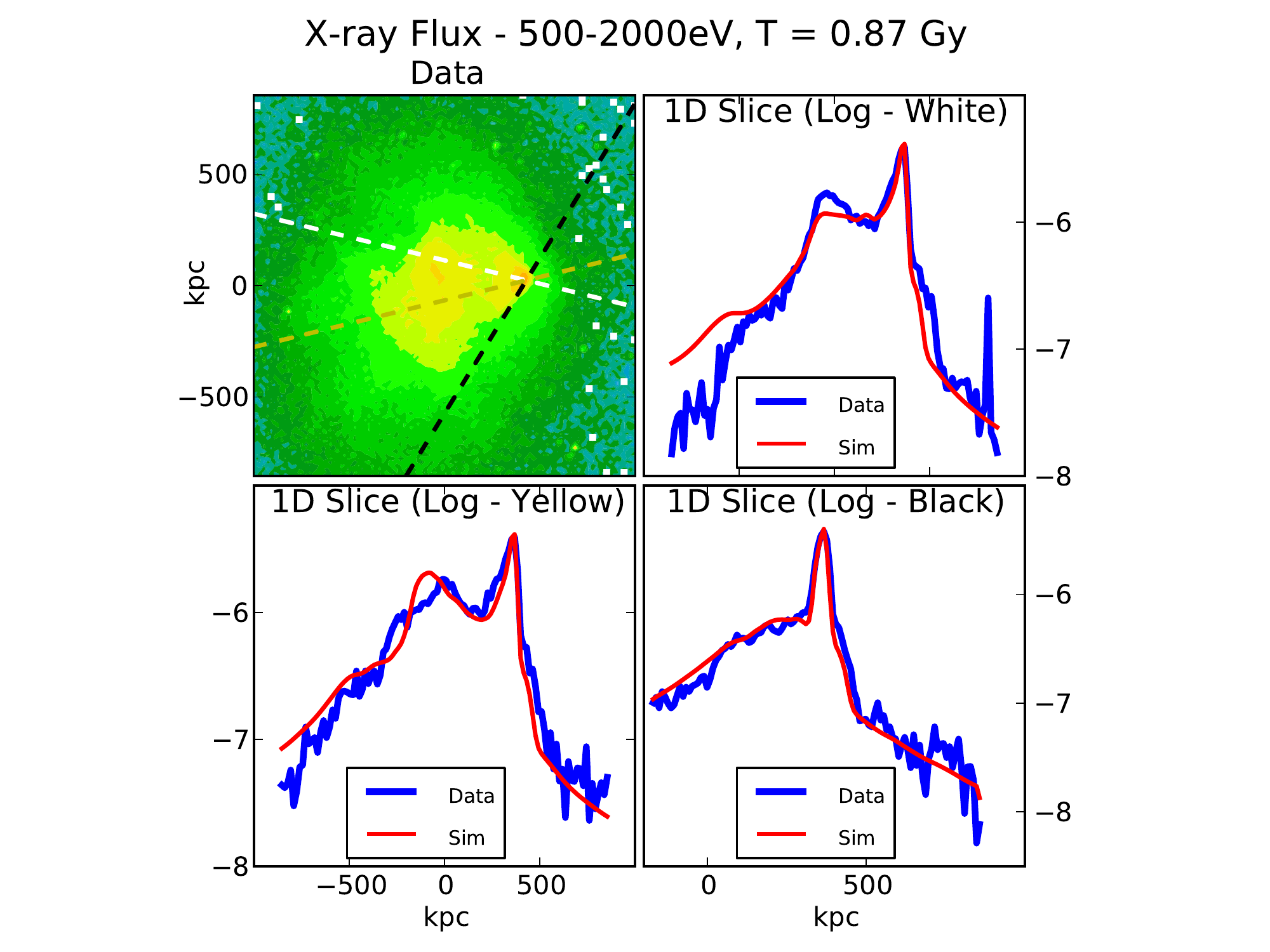}\label{Best_Fit_X-ray_Log}}
	\subfigure[500-2000eV Edge detection]{\includegraphics[trim = 1.0in 0.0in 0.7in 0.34in, clip, width=0.48\textwidth]{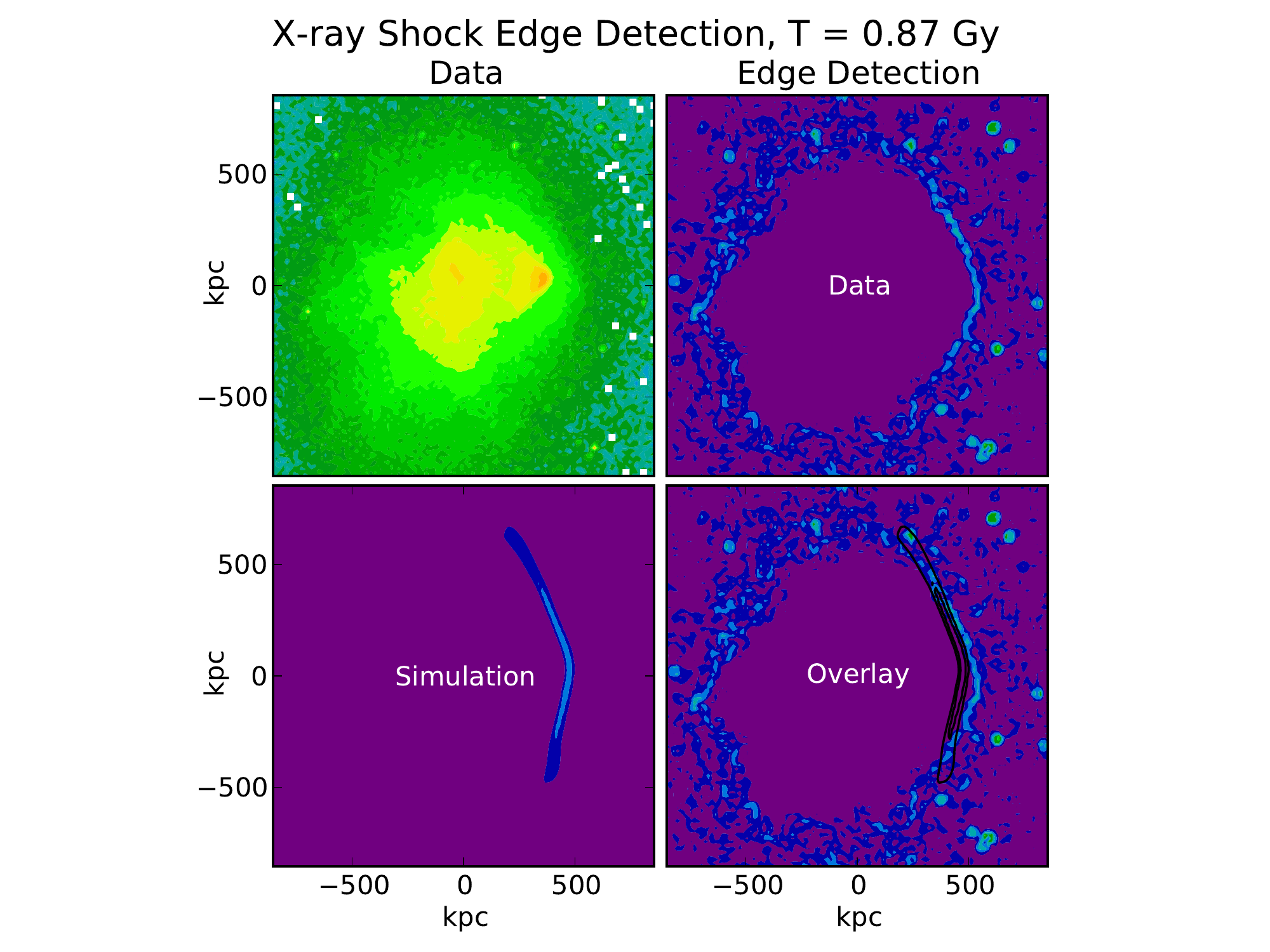}\label{Best_Fit_X-ray_Shock}}
	\caption{Best fit result for the X-ray flux in the range of 500-2000 eV.  The left hand plot has slices on a log scale at three different angles. The right hand plot uses an edge detection algorithm to capture the location of the shock. These two plots show that the shock location and shape are reasonably well captured.}
\end{figure}

%\begin{it}X-ray Flux:\end{it} \\
\label{XrayFlux}
Simultaneously fitting the mass lensing data and the X-ray flux data is more difficult than fitting the mass lensing alone, which is not surprising given the complexity of the baryonic physics and possible systematic errors in the mass lensing reconstruction.   Figures \ref{Best_Fit_X-ray1} - \ref{Best_Fit_X-ray3} show the fit to the X-ray fluxes in the three different energy bins.  Figure \ref{Best_Fit_X-ray_Log} shows the X-ray flux from different slices through the system, on a log scale, and Figure \ref{Best_Fit_X-ray_Shock} shows the location of the shock; these two plots are intended to show how well we have captured the location and shape of the shock.  The fit is reasonable, and in particular Figure \ref{Best_Fit_X-ray_Log} shows that the X-ray flux is well modelled over more than 2 orders of magnitude.  However, the $\chi^2$ calculated only from the lowest energy X-ray data has a value of 5.68, so this fit is not nearly as good as the mass lensing fit.

\begin {figure}[H]
	\centering
	\subfigure[Baryon profiles of the two clusters; `main' cluster on the left, `bullet' cluster on the right.]
        {\includegraphics[trim=0.3in -0.3in 0.4in 0.2in,clip,width=0.48\textwidth]{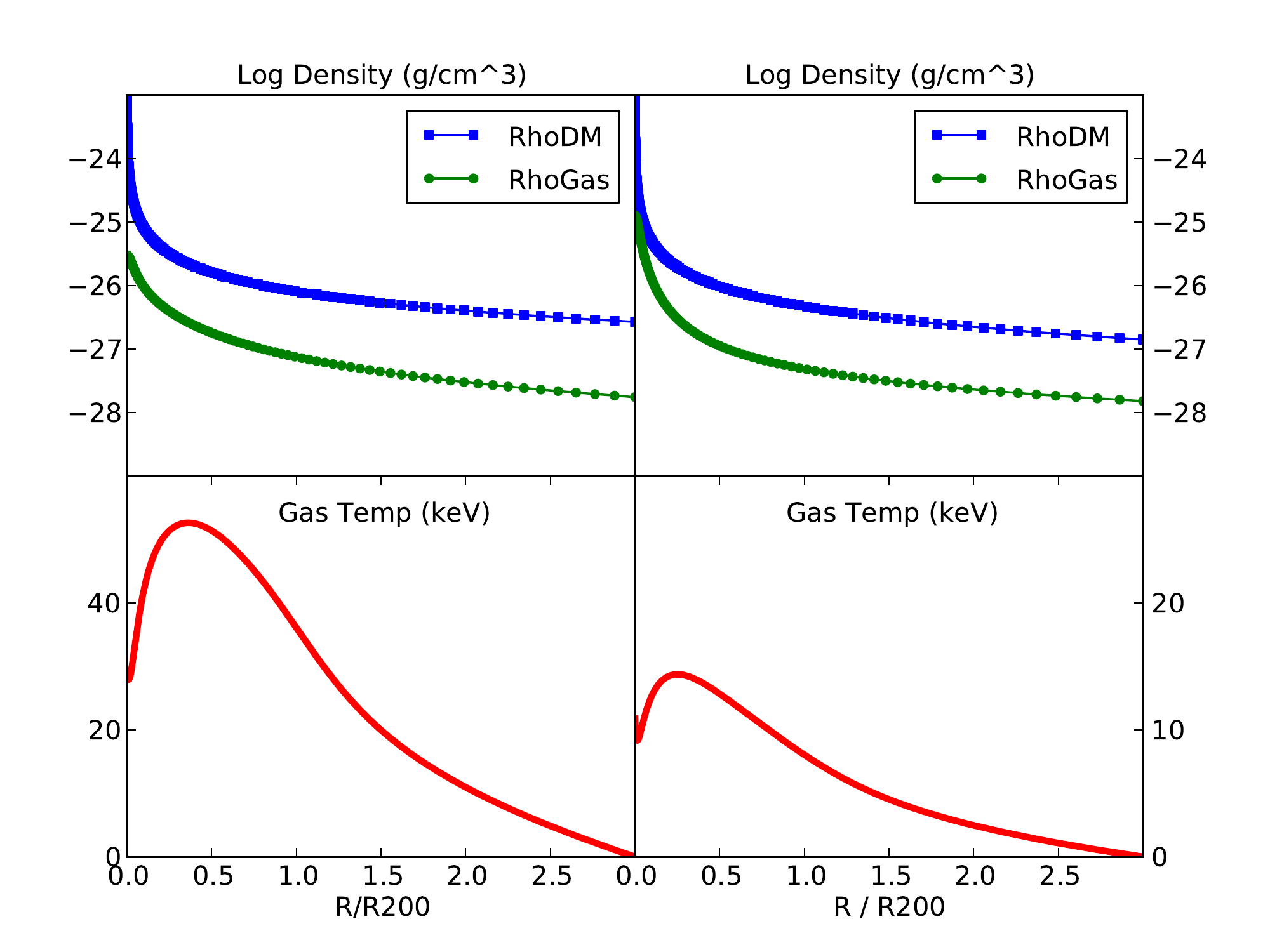}\label{Initial_Cluster_Profiles}}
  % Trim is Left Bottom Right Top
  % These plots are created from the mainprofile.dat and bulletprofile.dat files (which are created by the TriaxialHalo_SGas program)
  % using the program /bullet/code/pysubs/plotprofiles_sgas_25Jun13.py. This creates a file called Graph_Profiles_Paper.pdf.
  % These plots are from bsearch_matrixn110/batchfiles/batch3/ddfiles/run40/Graph_Profiles_Paper.pdf
	\subfigure[Temperature profiles of measured clusters.]
        {\includegraphics[trim=1.9in 0.8in 1.9in 1.5in,clip,width=0.48\textwidth]{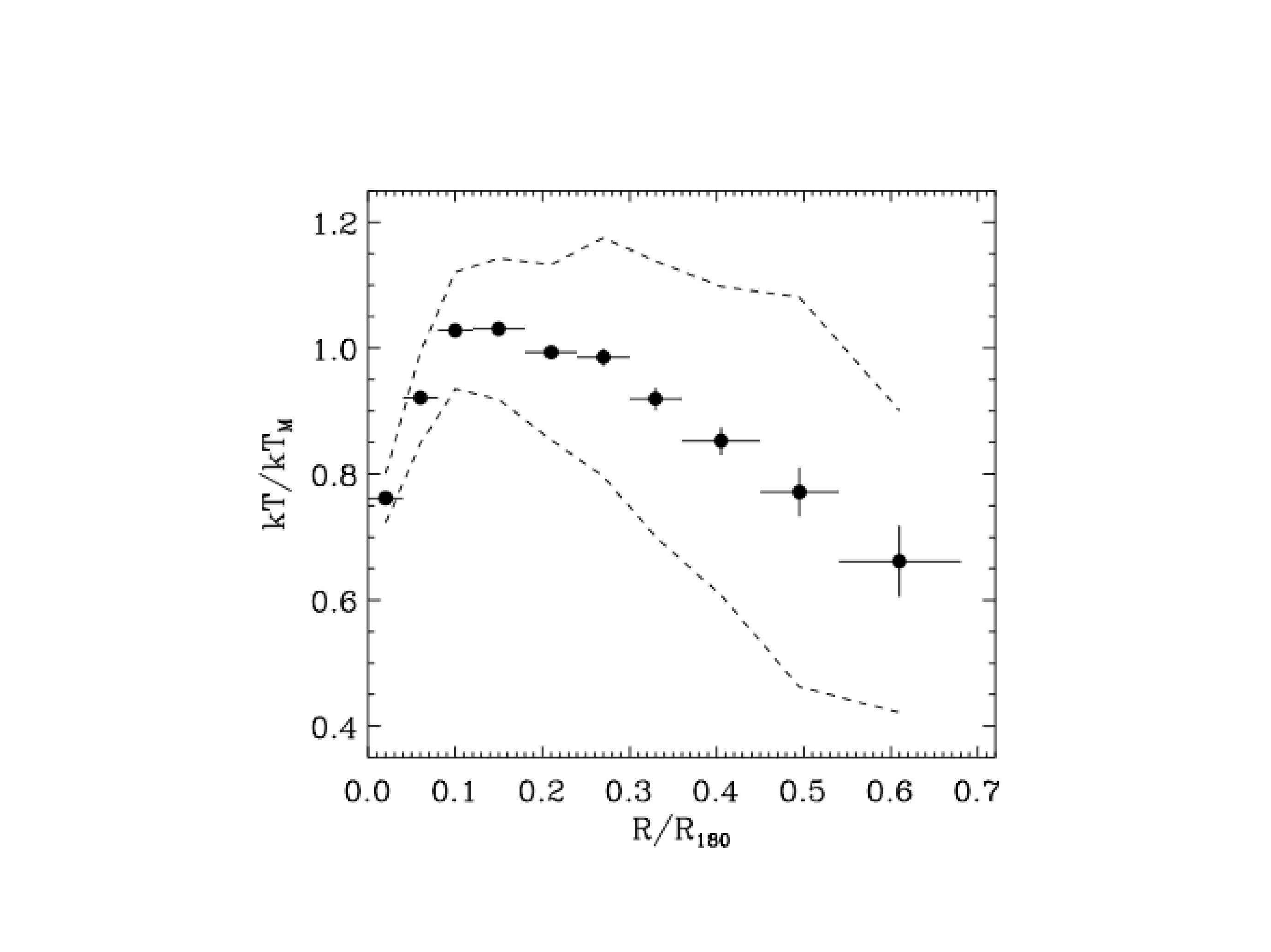}\label{Temp_Comparison}}
	\caption{Initial best-fit baryon profiles of the two clusters, and comparison of our initial temperature profiles to a large sample of X-ray clusters.  Figure (b) is reproduced from Leccardi and Molendi \cite{leccardi2008}; black points are the sample mean and the dotted lines are the one $\sigma$ scatter.  Our initial temperature profiles are qualitatively similar.}
\end{figure}

The X-ray flux morphologies are found to be quite sensitive to the details of the initial baryon distributions, and the initial distributions which give the best fit are shown in Figure \ref{Initial_Cluster_Profiles}.  It is important to note that the temperatures plotted here are effective temperatures including the effects of non-thermal pressure, and are therefore higher than the true temperatures, as discussed in Sections \ref{Non-Thermal_Pressure_1} and \ref{Non-Thermal_Pressure_2}.  To see whether our profiles are reasonable, we turn to the extensive measurements of single galaxy clusters which have been done, including McCourt et.al. \cite{mccourt2012}, Leccardi and Molendi \cite{leccardi2008}, and Simionescu et.al. \cite{simionescu2011}.  Figure \ref{Temp_Comparison} shows measured results of cluster temperature profiles from Leccardi and Molendi, which are seen to be qualitatively similar to our initial temperature profiles. 

\subsubsection{S-Z Effect, Plasma Temperature, and Role of Non-Thermal Pressure}
\label{Non-Thermal_Pressure_2}
\begin {figure}[H]
	\centering
	\subfigure[With non-thermal pressure.]{\includegraphics[trim = 1.0in 0.0in 0.7in 0.34in, clip, width=0.48\textwidth]{Best_Fit02.pdf}}
	\subfigure[No non-thermal pressure.]{\includegraphics[trim = 1.0in 0.0in 0.7in 0.34in, clip, width=0.48\textwidth]{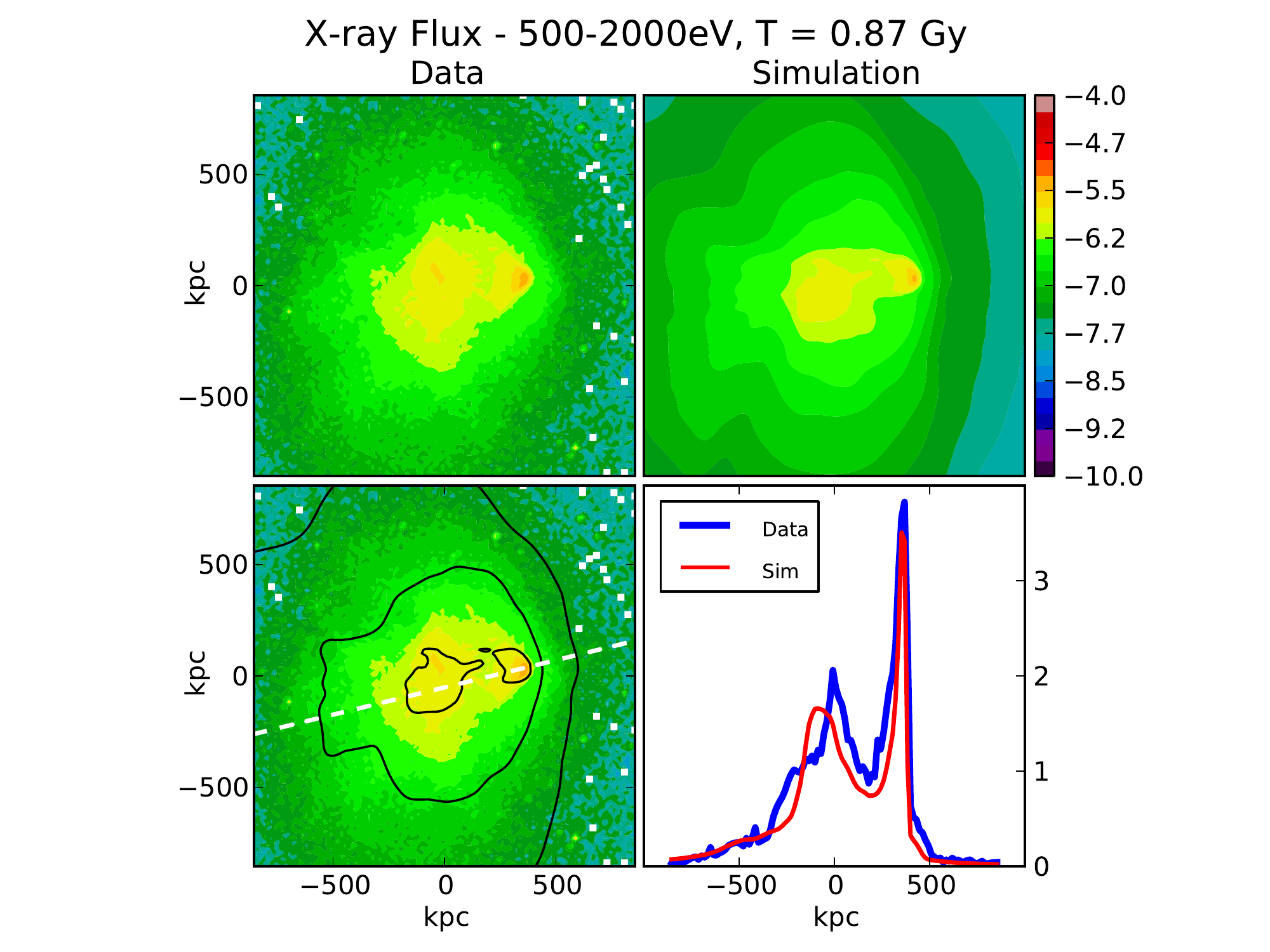}}
  \caption {Impact of the non-thermal pressure on the X-ray flux from 500-2000 eV.}
  \label{Non_Thermal_Pressure_02}
  % trim is L B R T
  %These are the standard plots from bsearch_matrixn110/batchfiles/batch3/ddfiles/run40 with TFudge set to 1.0
  \end{figure}

As discussed in Section \ref{Non-Thermal_Pressure_1}, we find it necessary to explicitly include a fudge-factor to account for effects of non-thermal pressure in order to correctly describe some datasets.  We describe the non-thermal pressure with a single parameter $\rm f_{ntp}$ which we take to be constant in space and time.  Not including non-thermal pressure in the simulation has a minor impact on the X-ray flux in the 0.5-2 keV band used to constrain the parameters (see Figure \ref{Non_Thermal_Pressure_02}), but leads to temperature and S-Z effect comparisons which are far out of agreement with the observations.  

\begin {figure}[H] 
	\centering
	\subfigure[With non-thermal pressure.]{\includegraphics[trim = 1.0in 0.0in 0.7in 0.34in, clip, width=0.48\textwidth]{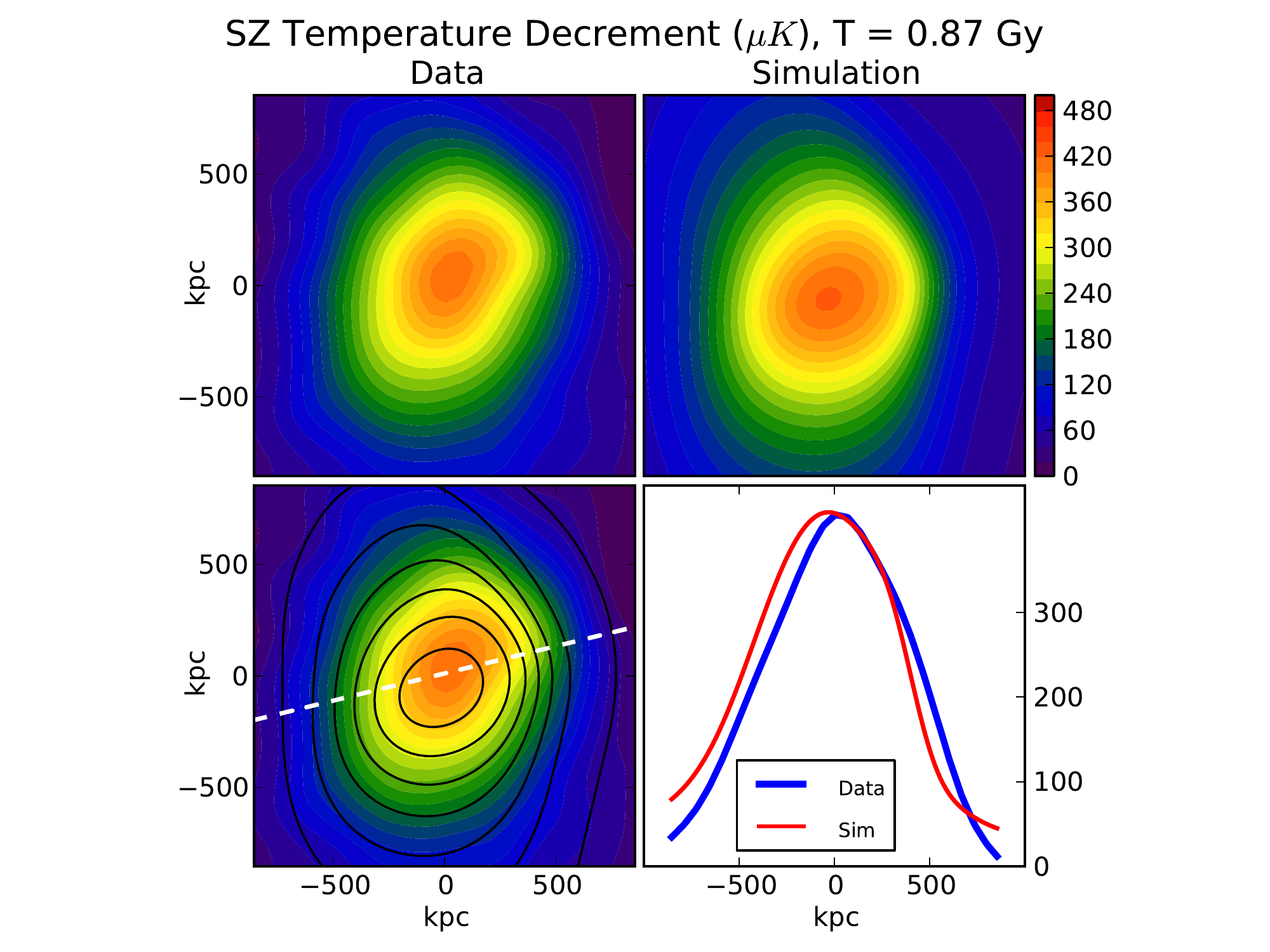}\label{Best_Fit_SZE}}
	\subfigure[No non-thermal pressure.]{\includegraphics[trim = 1.0in 0.0in 0.7in 0.34in, clip, width=0.48\textwidth]{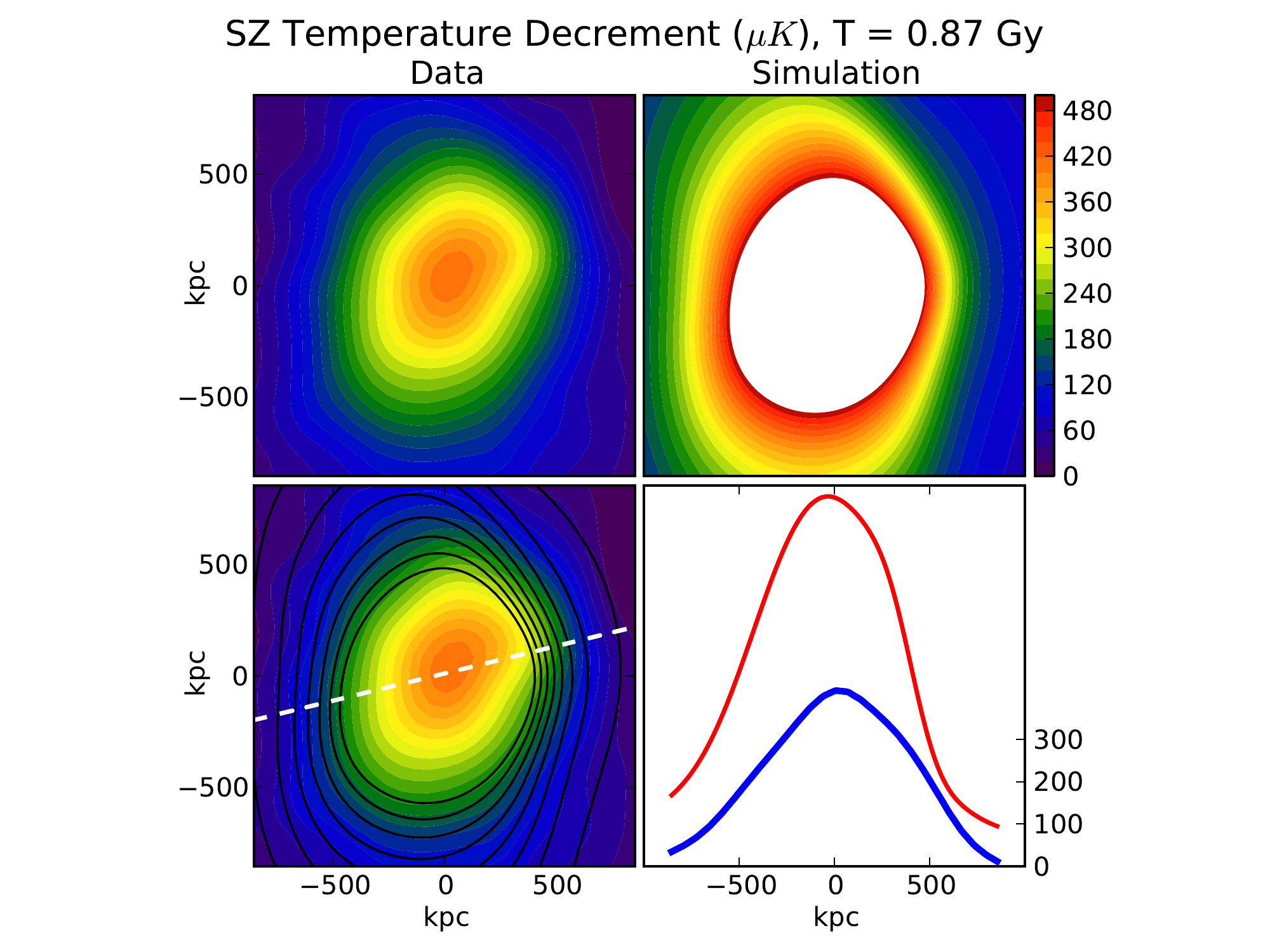}\label{Non_Thermal_Pressure_08}}
  \caption {Fit results for the S-Z temperature decrement, with and without the impact of non-thermal pressure.  The contour plot color saturation in Figure (b) results from ensuring all plots are on the same scale.}
  \end{figure}

Figure \ref{Best_Fit_SZE} shows the S-Z effect data.  Its sensitivity to the non-thermal pressure is shown in Figure \ref{Non_Thermal_Pressure_08}. The overall structure of the SZ observations is well-matched, but clearly the normalization cannot be predicted with accuracy until non-thermal pressure is treated better.  The offset between predicted and observed emission peaks is similar to the net offset between predicted and observed peaks in X-ray and mass lensing peaks;  whether any significance can be attached to that is under study. 

\begin {figure}[H]
	\centering
	\subfigure[With non-thermal pressure.]{\includegraphics[trim = 1.0in 0.0in 0.7in 0.34in, clip, width=0.48\textwidth]{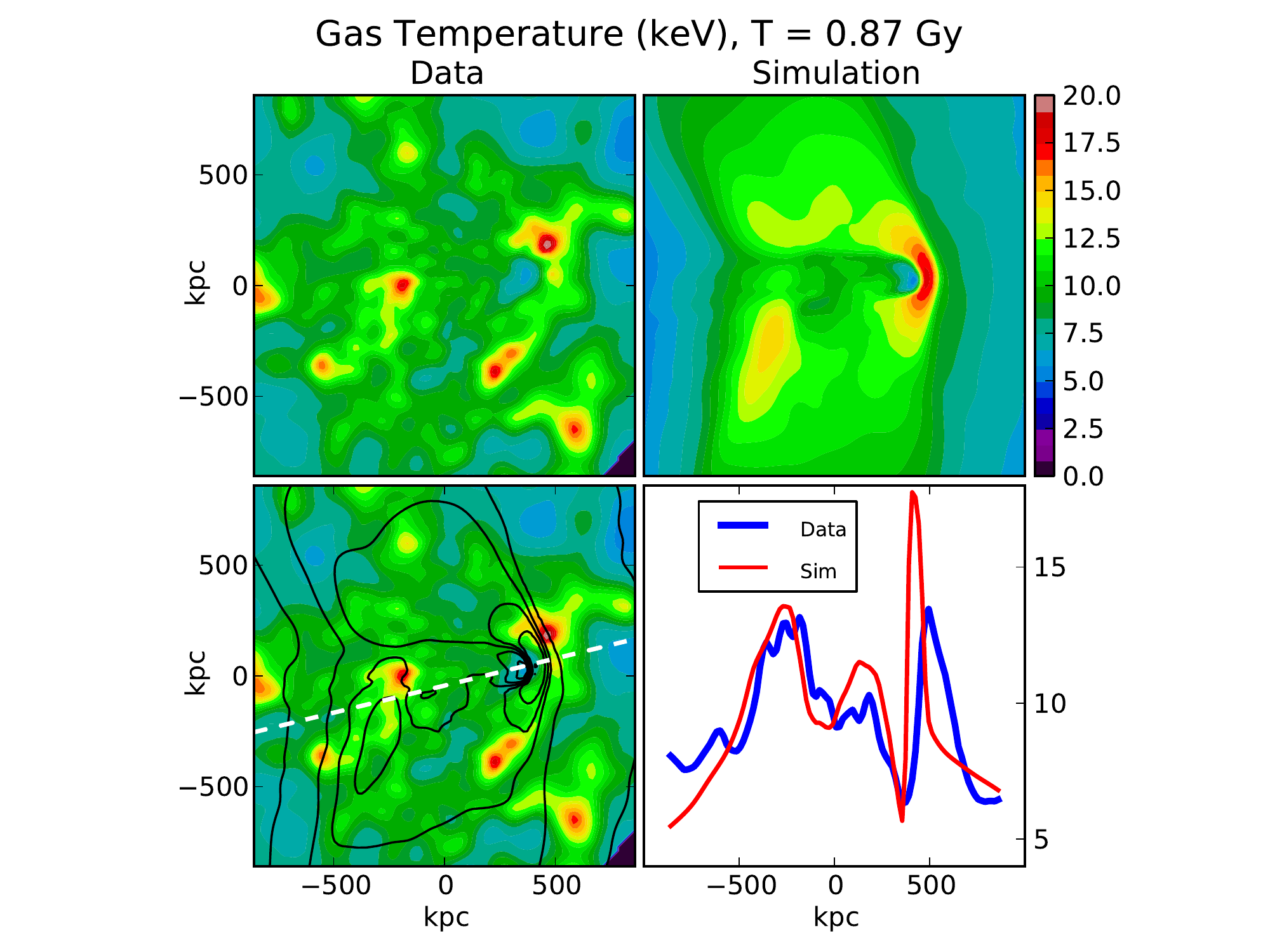}\label{Best_Fit_Temp}}
	\subfigure[No non-thermal pressure.]{\includegraphics[trim = 1.0in 0.0in 0.7in 0.34in, clip, width=0.48\textwidth]{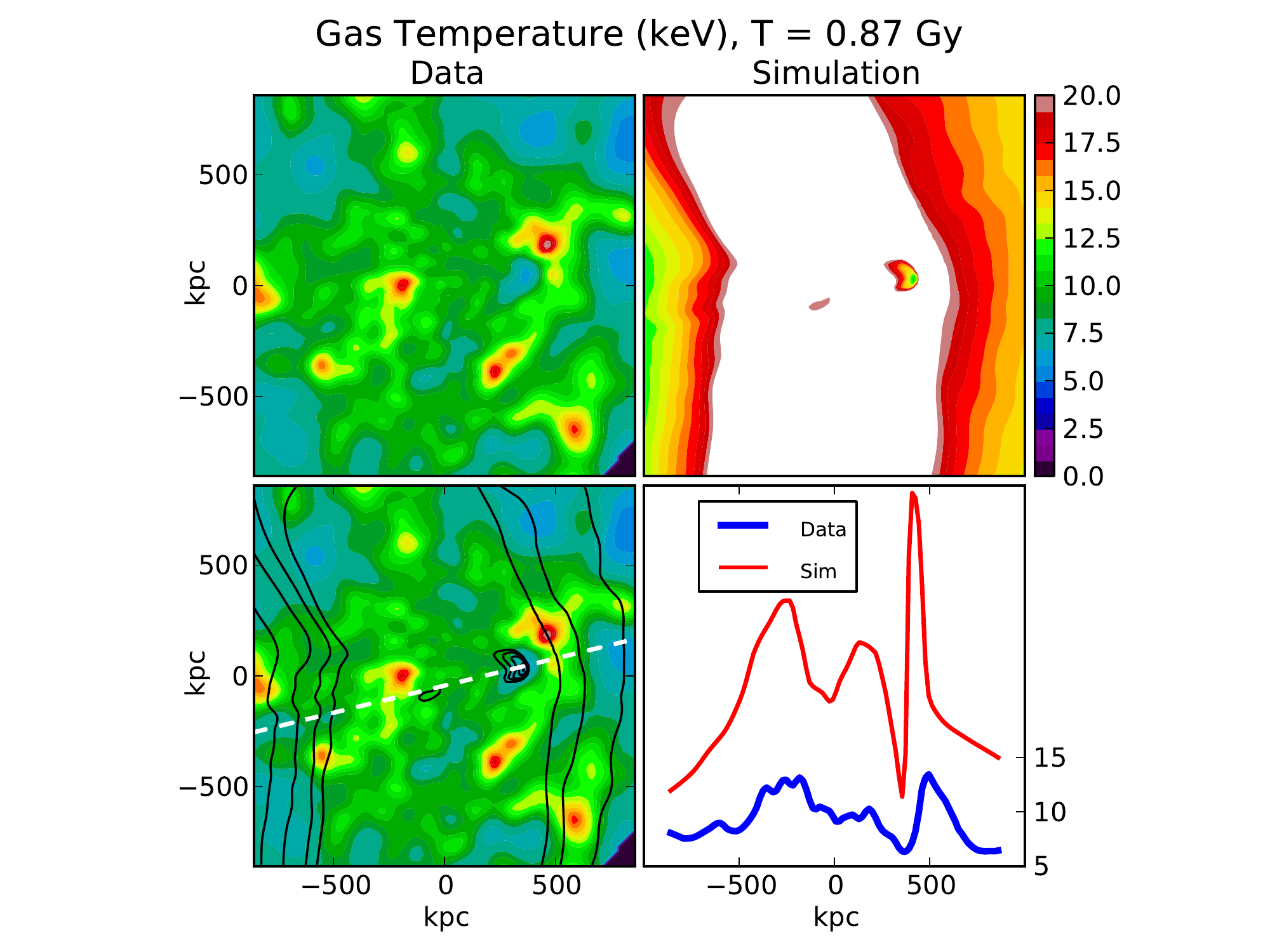}\label{Non_Thermal_Pressure_07}}
  \caption {Fit results for the plasma temperature with and without the impact of non-thermal pressure.  The contour plot color saturation in Figure (b) results from ensuring all plots are on the same scale.}
  \end{figure}

Figure \ref{Best_Fit_Temp} shows the predicted plasma temperature averaged along the line of sight in each pixel, which is compared to the map extracted from the X-ray measurements by M. Markevitch using the procedure described in \cite{Markevitch_TMap} and kindly provided to us.  The temperature uncertainty is said to have a median value of 1.5 keV [private communication, 2011], but note that the extracted temperature along a given line of sight is susceptible to large variations due to Poisson statistics in the high-energy X-ray band with the result that the observational map contains a great deal of noise.  Therefore, only large-scale features should be compared to the predictions.  On the simulation side, non-thermal pressure is important for the high-energy X-ray band, and thus the overall scale of the temperature map is uncertain, as seen in Figure \ref{Non_Thermal_Pressure_07} showing the sensitivity of the temperature map to the non-thermal pressure.  Nonetheless, the qualitative features of hot, low density gas leading a dense cold core are clear in the predictions and also visible as a general pattern underlying the noise in the extracted temperature map.

Going forward, it is clear that treating the non-thermal pressure as a constant ratio relative to the thermal pressure is too simplistic and we plan to improve our treatment, as will be discussed in Section \ref{Discussion_Section}. 

\subsubsection{Role of Magnetic Field and Radio Halo Prediction}
\label{Mag_Field_2}
\begin {figure}[H]
	\centering
	\subfigure[Mag = 61$\mu G$.]{\includegraphics[trim = 1.0in 0.0in 0.7in 0.34in, clip, width=0.48\textwidth]{Best_Fit02.pdf}}
	\subfigure[Mag = 0.01$\mu G$.]{\includegraphics[trim = 1.0in 0.0in 0.7in 0.34in, clip, width=0.48\textwidth]{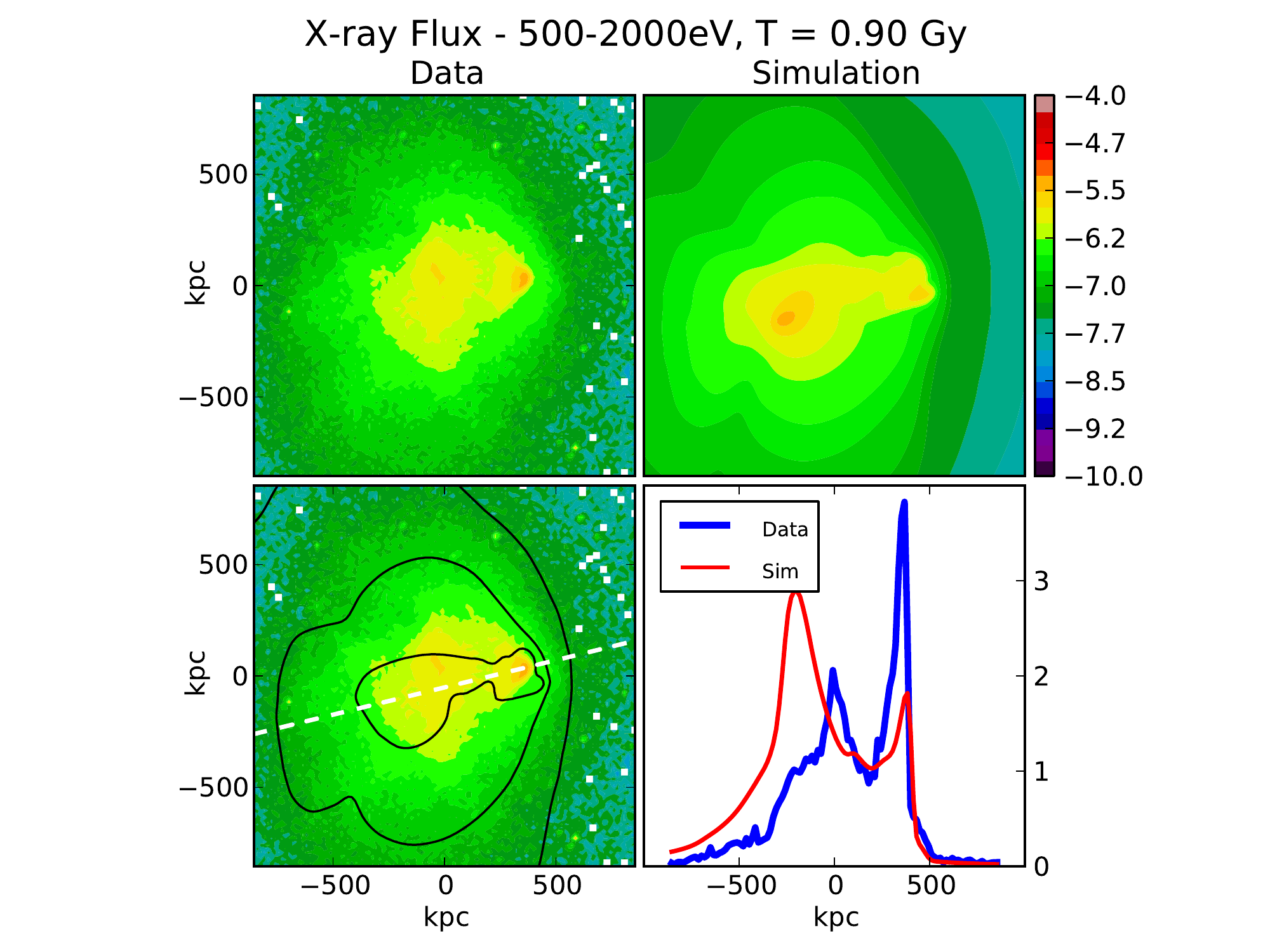}}
  \caption {Impact of magnetic field on the X-ray flux in the range of 500-2000 eV.  The magnetic field pressure impacts the location of the X-ray peaks.}
  \label{B_Impact}
 % These are the standard plots with B = 0.0
 % These are actually run in bsearch_matrixn110/batchfiles/batch3/ddfiles/run41
  \end{figure}

%\begin{it}Role of Magnetic Field and Radio Halo Prediction:\end{it} 
As noted above, it is non-trivial to accurately match the mass lensing reconstruction while simultaneously accurately reproducing the X-ray flux morphology.  The inclusion of magnetic fields is important for achieving a good description.  Figure \ref{B_Impact} shows that, given a good fit to the mass lensing distributions, the baryon density peaks (and their associated X-ray flux peaks) without the magnetic field tend to be ``ahead'' of their observed locations.  This requires some added pressure on the baryons in order to retard the motion of the baryon density peaks to agree with the observations.  We have achieved this with a combination of increased magnetic field and added viscosity, both of which retard the motion of the baryon peaks.  Attempts to achieve alignment of the various peaks with only the addition of the magnetic field are not successful, and the addition of the viscosity term is required.  Even with these components added, however, the reproduction of the shapes of the regions of high X-ray flux (see Figures \ref{Best_Fit_X-ray1} - \ref{Best_Fit_X-ray3}), while close, is still not completely accurate.  

\begin{figure}[H]
	\centering
	\includegraphics[angle=0, scale=0.6]{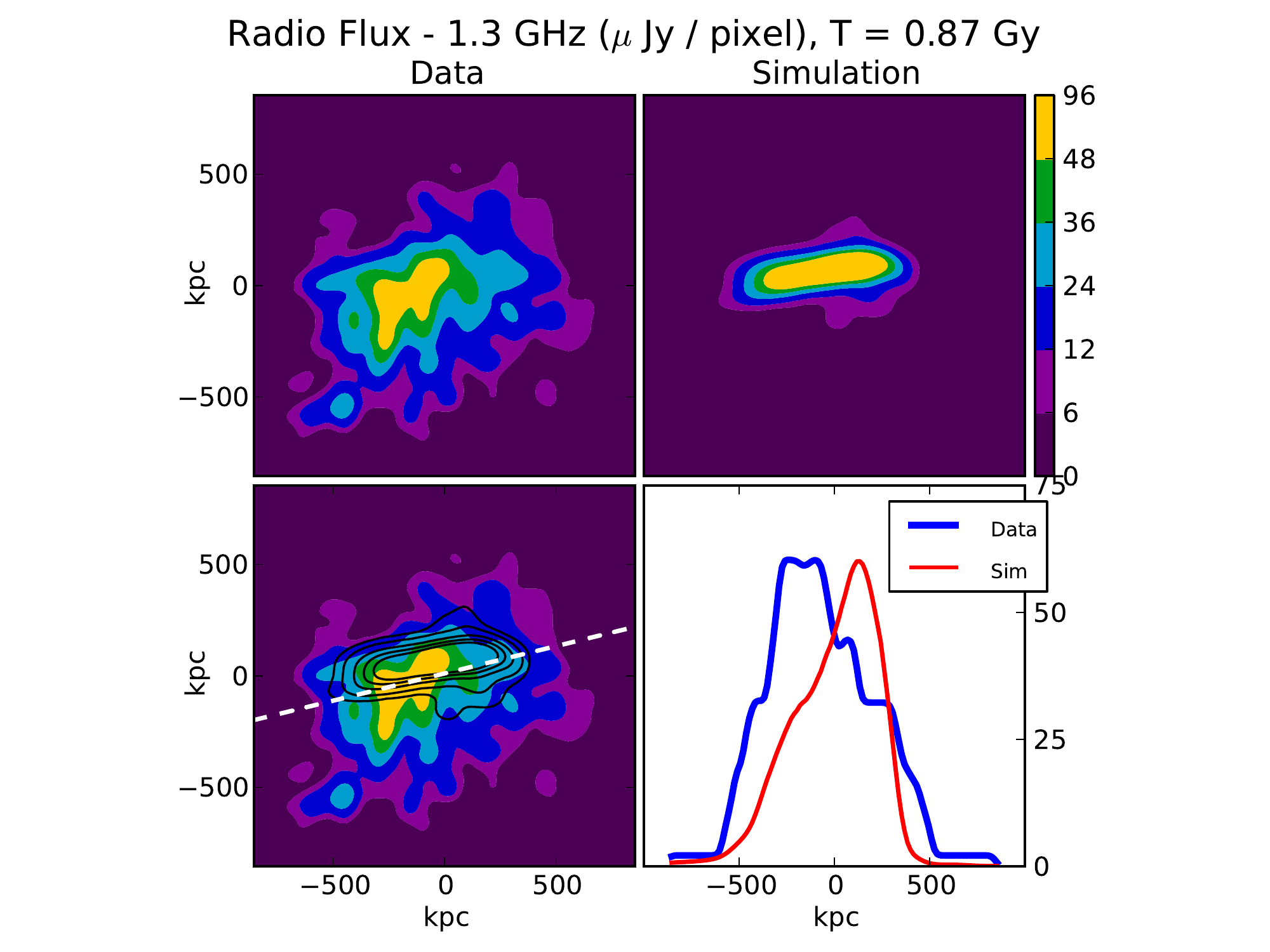}
	\caption{Predicted radio flux, as compared to observations by Liang, et. al.\cite{Liang}  The subpanels of each plot are as described in Figure \ref{Mass_only}.}
	\label{Radio_Plots}
 %This is from bsearch_matrixn110/batchfiles/batch3/ddfiles/run40 with P=3.6
  \end{figure}

\begin {figure}[H]
	\centering
	\subfigure[P = 3.4]{\includegraphics[trim = 4.02in 0.0in 1.0in 0.34in, clip, width=0.30\textwidth]{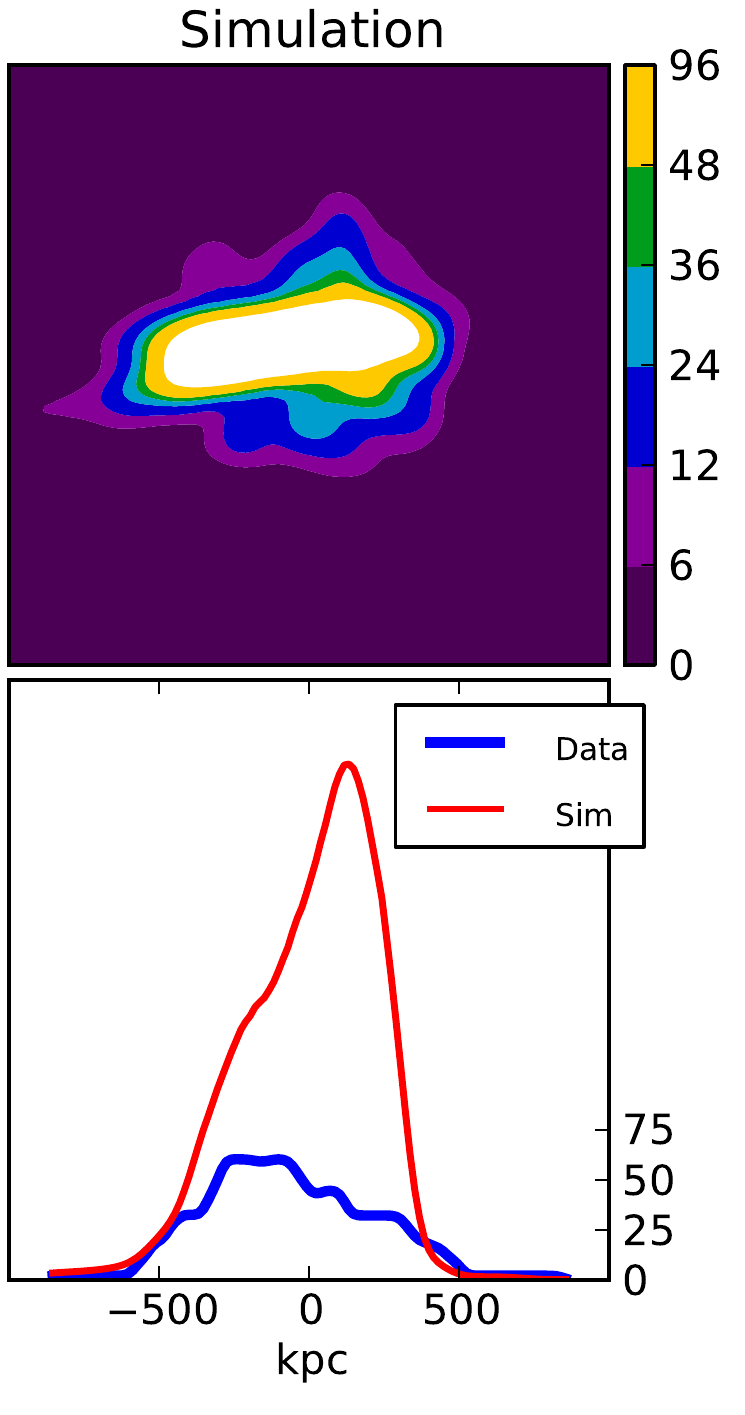}}
	\subfigure[P = 3.6]{\includegraphics[trim = 4.02in 0.0in 1.0in 0.34in, clip, width=0.30\textwidth]{Best_Fit09.pdf}}
	\subfigure[P = 3.8]{\includegraphics[trim = 4.02in 0.0in 1.0in 0.34in, clip, width=0.30\textwidth]{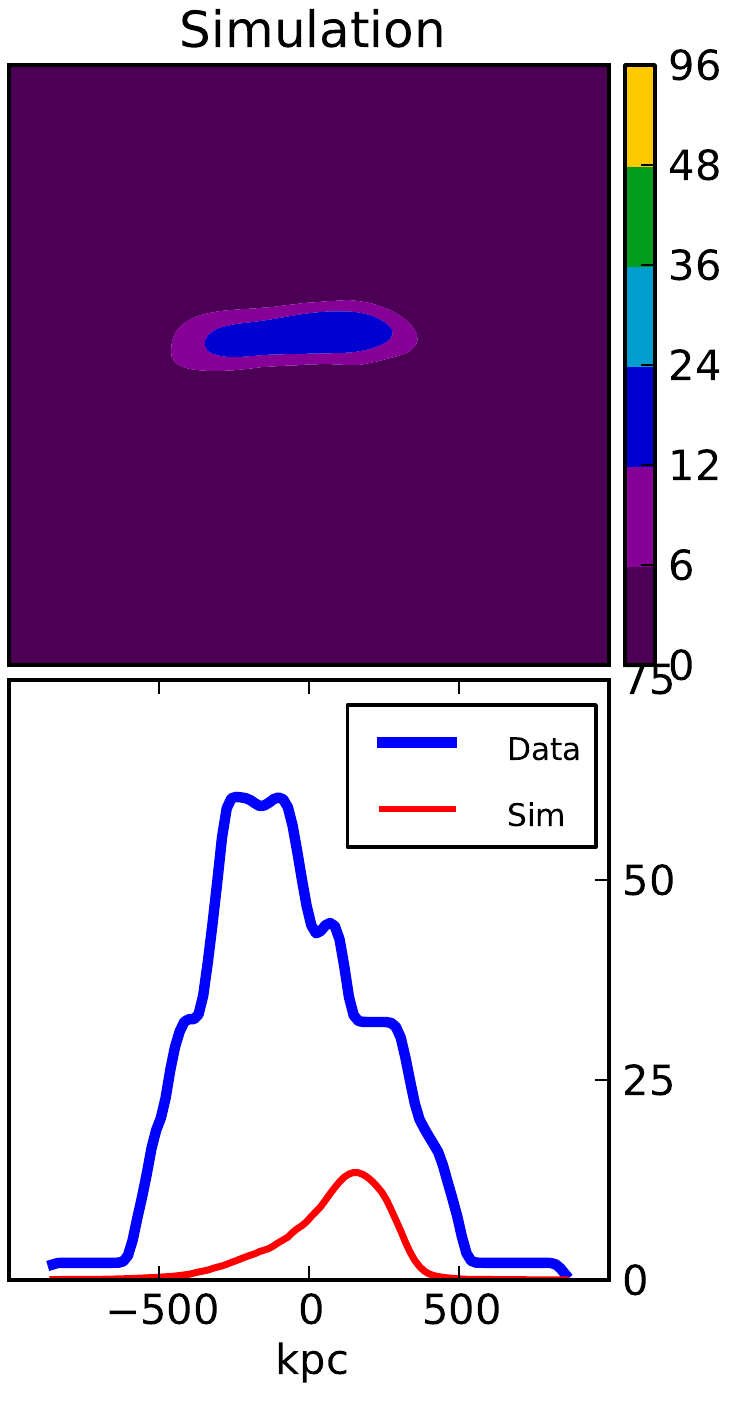}}
	% trim is L B R T
  \caption {Impact of the spectral index parameter $p$.  A value of $p$=3.6 best captures the magnitude of the radio halo flux.  The contour plot color saturation in Figure (a) results from ensuring all plots are on the same scale.}
  \label{Spectral_Index}
 %These are the standard plots from bsearch_matrixn110/batchfiles/batch3/ddfiles/run40 with P adjusted
  \end{figure}

\begin {figure}[H]
	\centering
	\subfigure[Near the beginning of the simulation.]
		  {\includegraphics[trim=1.5in 0.2in 2.62in 0.2in, clip, width=0.48\textwidth]{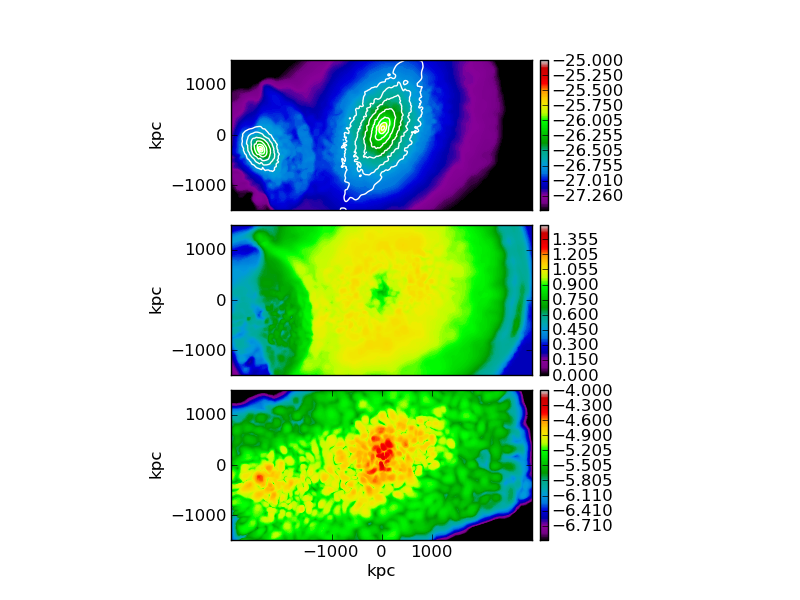}}
	\subfigure[At the time of observation.]
		  {\includegraphics[trim=2.26in 0.2in 1.86in 0.2in, clip, width=0.48\textwidth]{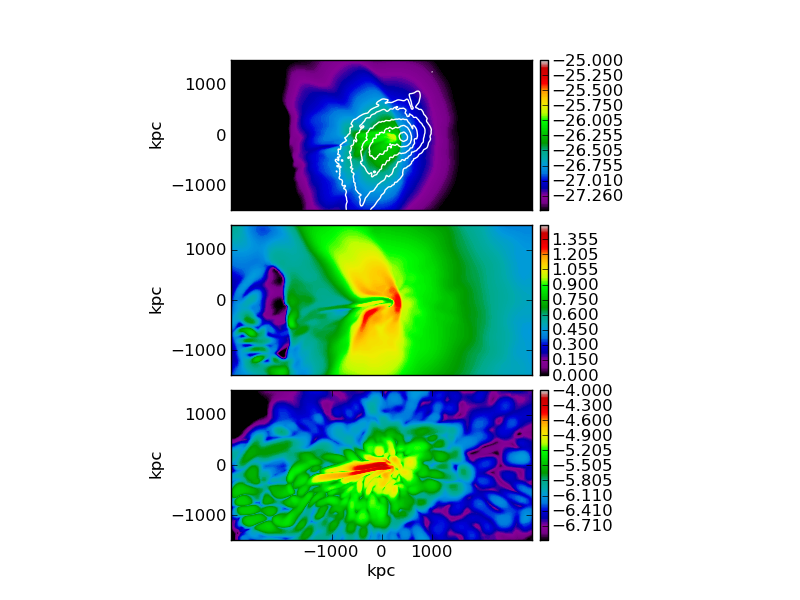}}
		  \vspace{-0.15in}
  \caption {Top panel: log(baryon density in $\rm g/cm^3$) plotted in color; dark matter density overlaid in white contours. Middle panel: log(gas temperature in keV). Bottom panel: log($\rm|B|$) in Gauss.}
  \label{Initial_and_Final_Slices}
  % These are created in the appropriate directory using the program /bullet/code/pysubs/movie.py
  % These are from bsearch_matrixn110/batchfiles/batch3/ddfiles/run4/movie/collision012.png and collision088.png
  % trim is L B R T
  \end{figure}

While the best-fit magnetic field found here does not play a dominant dynamical role, the details of the initial magnetic field distribution (see Section \ref{Mag_Field_1}) do impact the X-ray flux results.  To further explore whether the magnetic fields found here are reasonable, we calculated the radio halo emission, and compared this to the measurements of Liang et.al. \cite{Liang}.  As detailed in Appendix \ref{Radio_Calculation}, we use a simple model where a population of relativistic electrons is taken to be in equipartition with the magnetic field.  This population follows a power law distribution with power law exponent $p$ (see Equation \ref{ESpectrum}), and produces a radio flux with spectral index $s$ (see Equation \ref{SIndex}), where $p$ and $s$ are related by Equation \ref{P_s_Equation}.  Figure \ref{Radio_Plots} shows the fit to the radio halo data using a typical magnetic field as determined from the collision dynamics; it is to be emphasized that since the initial B-field is randomly generated, a detailed fit is not the goal: reproducing the general magnitude and location of the radio emissions is the best that can be expected.   The value of the power law exponent $p$ is fairly tightly constrained to a value of $ p \sim3.6$, as seen in Figure \ref{Spectral_Index}, which shows that this value is needed in order to reproduce the magnitude of the observed radio emissions.  This predicts a value of radio emission spectral index $s\sim1.3$, which is encouragingly close to the value of 1.2-1.3 measured by Liang.  The fact that the value of magnetic field which is required to give the proper alignment of the X-ray intensity peaks is consistent with the radio halo emission lends confidence to the model.  Future work will explore this further to see if modifications of the initial B-field or improvements to the radio emission model (Appendix \ref{Radio_Calculation}) can improve this fit further.  Figure \ref{Initial_and_Final_Slices} shows the magnetic field amplification which takes place during the collision.

% Impact of Spectral Index 

\section{Discussion}
\label{Discussion_Section}
%\subsection{Comparison to Universal Baryon Fraction}
The fit between the observations and the simulation, while far from perfect, is remarkably good.  The physics incorporated into the simulations is conventional, and the assumed initial conditions are generally quite reasonable as compared to known galaxy clusters.  The best-fit baryon fraction values (within $\rm R_{200}$) of the initial clusters (parameters GF1 and GF2 in Table \ref{Fitting_Parameters} - best fit values of 19$\pm 2 \%$ and 17$\pm 2 \%$) are close to the $\rm \Lambda CDM$ average value.  The nine-year WMAP results \cite{WMAP2013}, for example, give a ratio of $\rm \Omega_b / \Omega_m$ of 16.5 $\pm 2.5 \%$, while the recent Planck results \cite{Planck2013} give a ratio of $\rm \Omega_b / \Omega_m$ of 15.4 $\pm 0.5 \%$.  Also, the metallicity Z is larger than is physical.  Further work is needed to determine whether the slightly higher baryon fraction values we find are due to a deficiency of our modeling, such as missing sub-grid physics, or whether the baryon fraction in these clusters is actually enhanced over the universe-average values.  The prime driver in the overall gas fraction normalizations is the X-ray luminosity, which $\sim n^{2}$, so including expected small-scale inhomogenieties would plausibly result in lower GFs as well as decrease the extracted metallicity. In the next phase of the work, a variety of such effects will be explored.   

%\subsection{Comparison to Past Simulation Studies}
We can compare the quality of the fit achieved here to that of the earlier simulation studies of Springel and Farrar \cite{sf07} and Mastropietro and Burkert \cite{Mastropietro}.   %.  While the simulation tool used here is different, and we do not have all of the details of these other simulations, we were able to run comparisons 
As noted in the Introduction, those works took a different approach of trying to fit some key separations between features, and did not explore such a large range of initial conditions as we have done.  Nonetheless, we can use the initial conditions reported in those papers in our simulation, and compare to the observations using our techniques.  The value of the figure-of-merit parameter $\rm \chi^2$ calculated from mass lensing and the lowest energy X-ray data is 3.92 for our best fit initial conditions, 13.67 for Springel and Farrar, and 19.93 for Mastropietro and Burkert.  This shows the clear improvement in fitting the data that we have achieved. 

  \begin{figure}[H]
	\centering
	\subfigure[This work]{\includegraphics[trim = 1.0in 0.0in 1.5in 0.34in, clip, width=0.331\textwidth]{Best_Fit01.pdf}}
	\subfigure[Springel and Farrar \cite{sf07}]{\includegraphics[trim = 1.58in 0.0in 1.5in 0.34in, clip, width=0.296\textwidth]{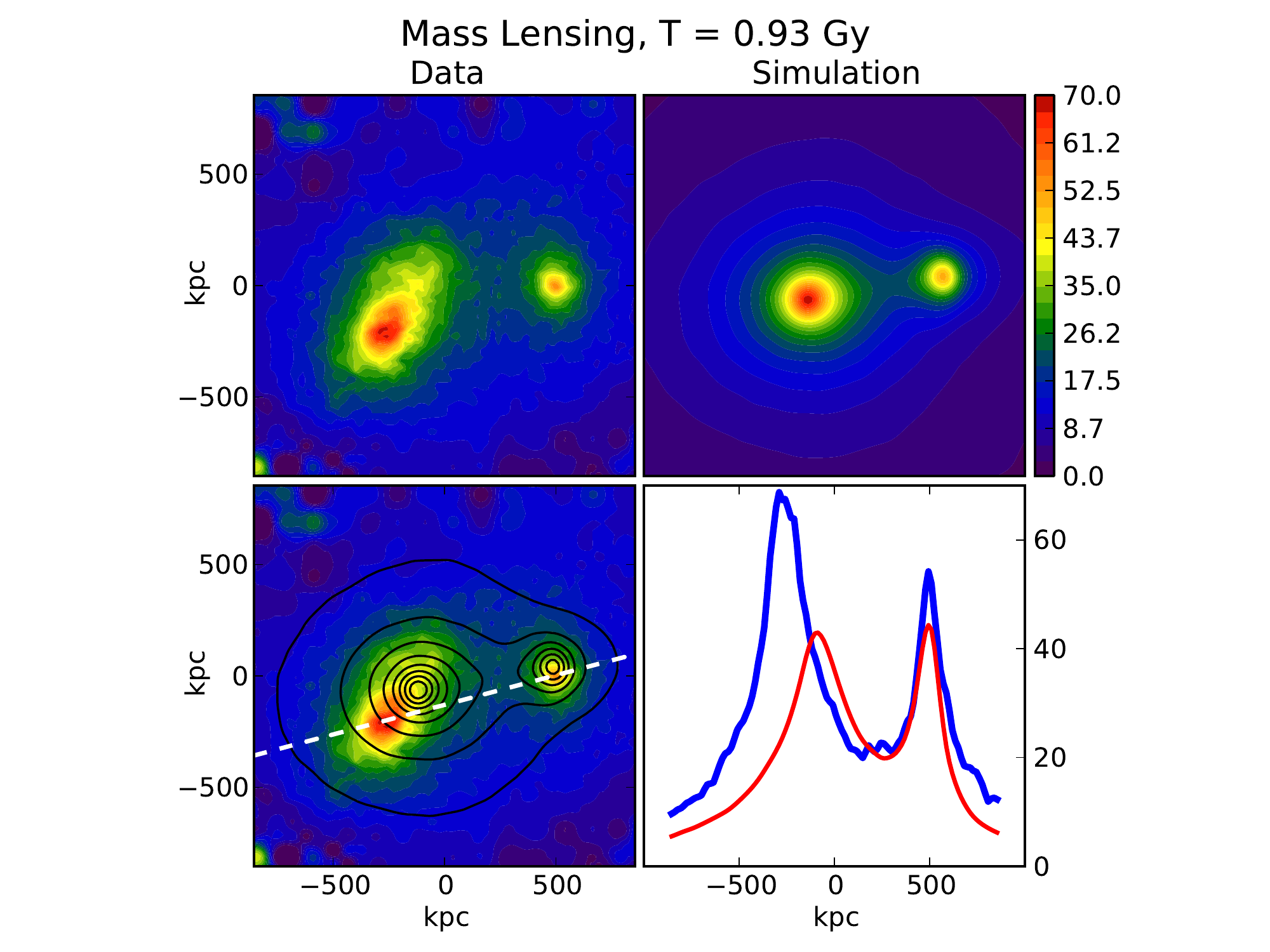}}
	\subfigure[Mastropietro and Burkert \cite{Mastropietro}]{\includegraphics[trim = 1.58in 0.0in 0.7in 0.34in, clip, width=0.344\textwidth]{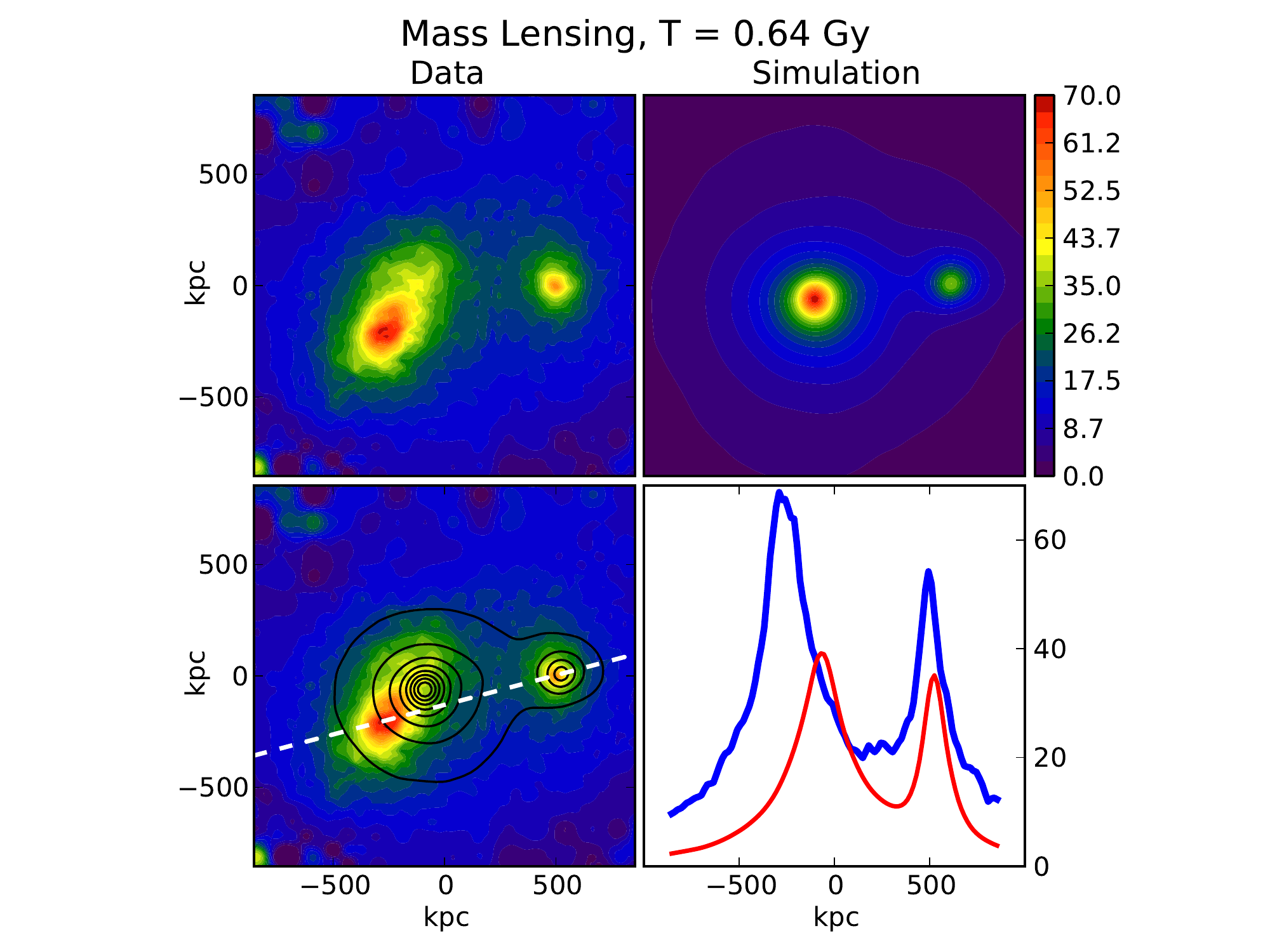}}
	\caption{Mass lensing fits as compared to past simulation work.  $\rm \chi^2$ calculated from mass lensing and lowest energy X-ray data as described in the text is 3.92 in this work, 13.67 for Springel and Farrar, and 19.93 for Mastropietro and Burkert.}
	\label{Mass_compare}
	% These are from our best case and mastropietro_burkert and springel_farrar
	% trim is L B R T
  \end{figure}

 \begin{figure}[H]
	\centering
	\subfigure[This work]{\includegraphics[trim = 1.0in 0.0in 1.5in 0.34in, clip, width=0.331\textwidth]{Best_Fit02.pdf}}
	\subfigure[Springel and Farrar \cite{sf07}]{\includegraphics[trim = 1.58in 0.0in 1.5in 0.34in, clip, width=0.296\textwidth]{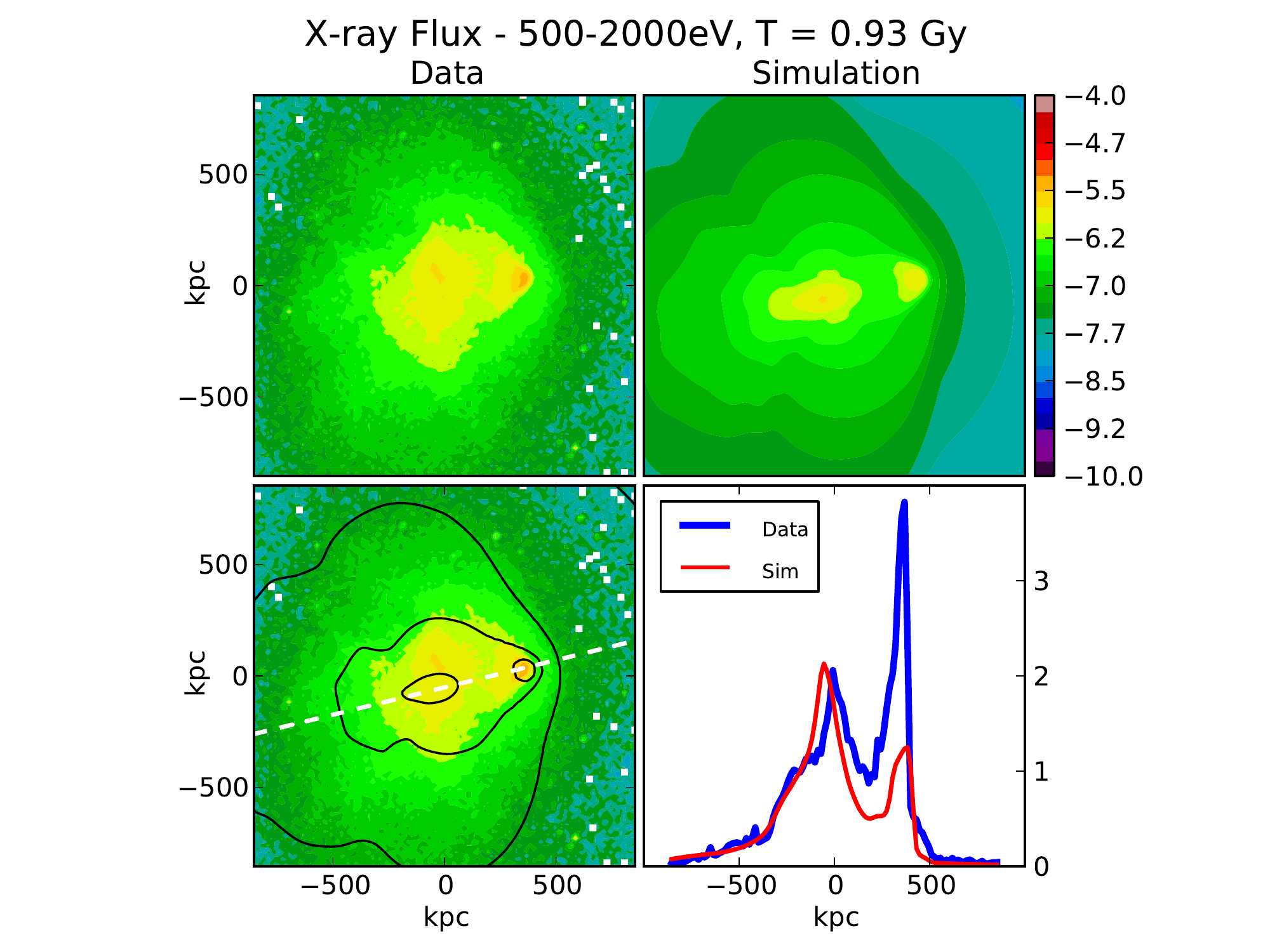}}
	\subfigure[Mastropietro and Burkert \cite{Mastropietro}]{\includegraphics[trim = 1.58in 0.0in 0.7in 0.34in, clip, width=0.344\textwidth]{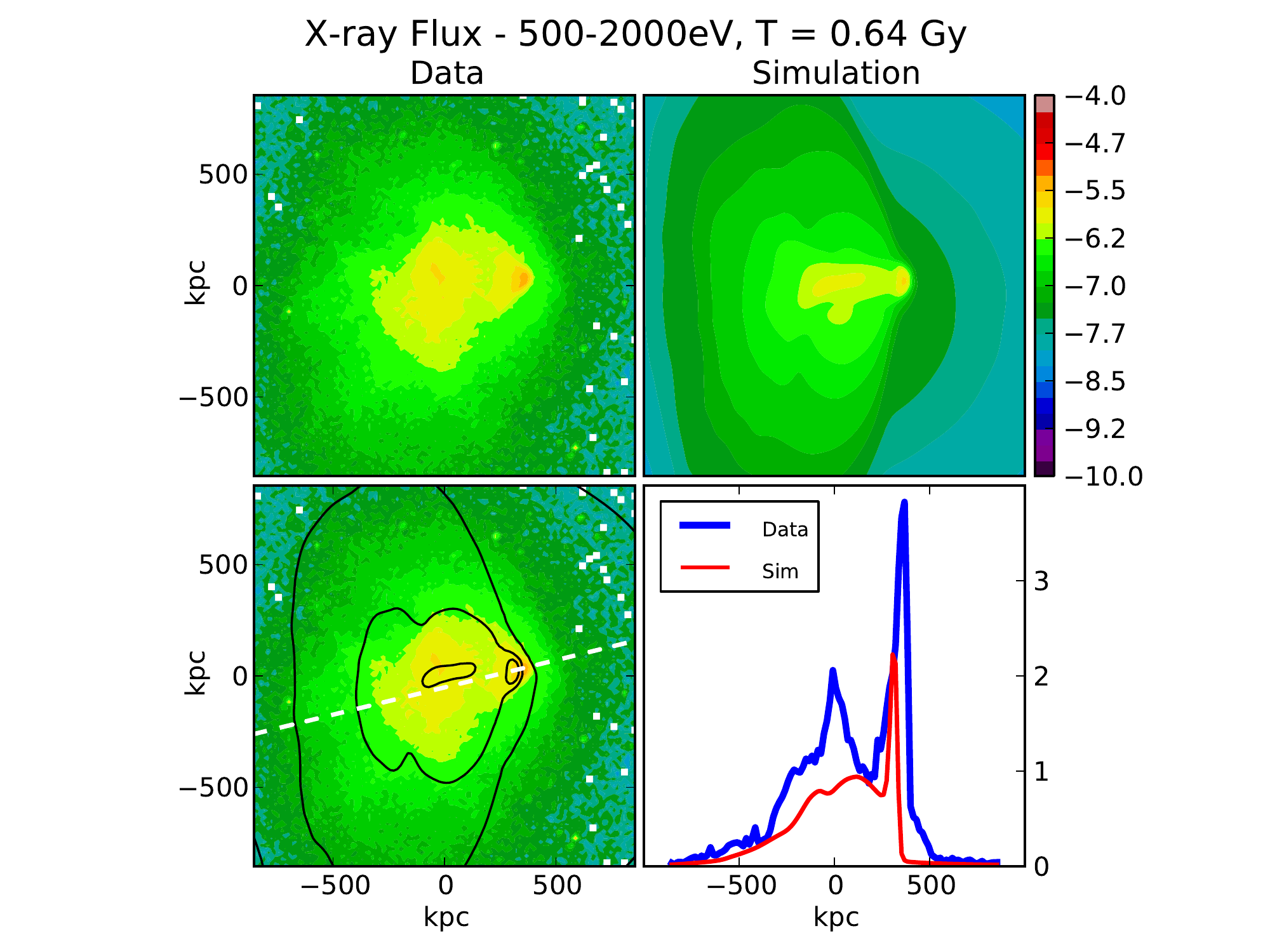}}
	\caption{Lowest energy X-ray fits as compared to past simulation work.  $\rm \chi^2$ calculated from mass lensing and lowest energy X-ray data as described in the text is 3.92 in this work, 13.67 for Springel and Farrar, and 19.93 for Mastropietro and Burkert.}
	\label{Xray1_compare}
	% These are from our best case and mastropietro_burkert and springel_farrar
	% trim is L B R T
  \end{figure}

\begin{figure}[H]
	\centering
	\includegraphics[trim=0.2in 1.4in 0.1in 0.5in,clip,width=\textwidth]{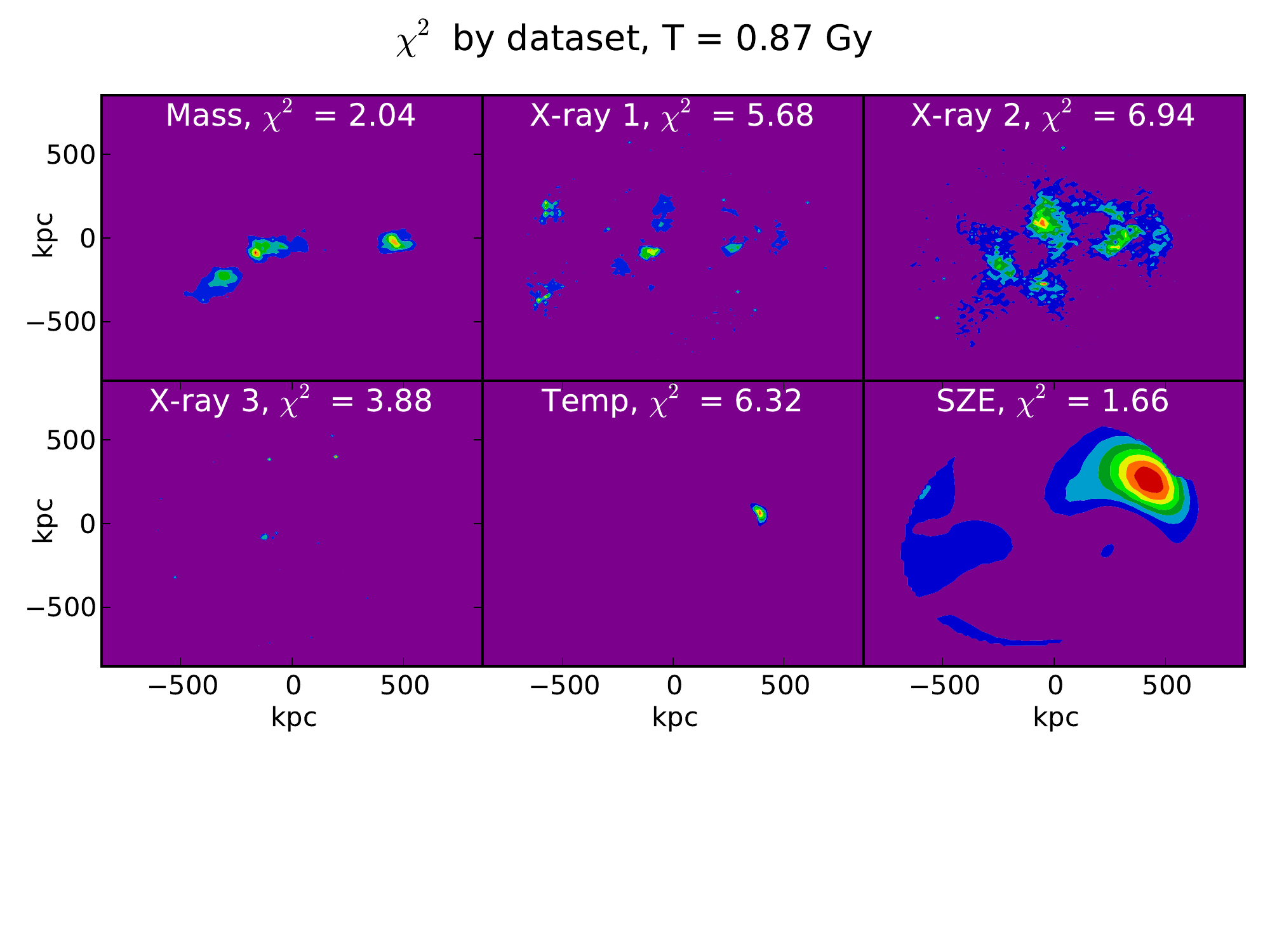}
	\caption{The $\chi^2$ value calculated for each dataset alone, with the location of the regions of poorest fit (i.e. the residuals) over the 2D simulation plane shown in the colored contours.}
	\label{Best_Fit_Chi_Squared}
	\end{figure}

Nonetheless, although the fits we have obtained are good and they are a considerable improvement over earlier attempts, it is clear that they could be improved.  Figure \ref{Best_Fit_Chi_Squared} shows the contributions to the total $\chi^2$ value (i.e. the residuals) over the 2D simulation plane, allowing the regions where the fit is poorest to be identified.  
Some directions of future development to improve the fidelity of this analysis are:
\begin{enumerate}
\item It is possible that we have simply not found the optimal initial conditions.  The parameter space is very large, and while multiple searches have been done, we have certainly not exhausted all possibilities.  Possibly more important, focusing simply on improving the $\chi^2$ as defined in Eq. \ref{Chi2_calculation} may be inadequate for producing a good fit to features like the shape of the shock, since this feature occurs in a region of relatively low X-ray flux.  We are exploring a more sophisticated figure of merit in order to better capture the shape of the shock region.  We remark, however, that the present definition of  $\chi^2$ already strikes a reasonably good balance between fitting the high-luminosity regions -- which are small but have low errors -- and large features at lower flux levels.  This is witnessed, for instance, by the good fit to X-ray flux over 2 orders of magnitude shown in Fig. \ref{Best_Fit_X-ray_Shock}.
\item The shape of the X-ray flux emission is sensitively dependent on the details of the baryon distribution.  The present analysis assumes that the baryons in the initial clusters are in hydrostatic equilibrium, so that the shape of the baryon density contours tracks the equipotential contours (see Section \ref{Baryon_Profiles}).  Lau, et.al. \cite{lau2011} have studied this in detail, and found that this approximation breaks down in the cluster central regions, where the dense gas can be rotationally supported in addition to being pressure supported.  Since it is these dense gas regions that contribute most heavily to the X-ray emissions, an enhancement along these lines could significantly improve the X-ray fits.  An effort along those lines is underway.  
\item Our treatment of non-thermal pressure can clearly be improved.  A phenomenological approach would be to take the non-thermal pressure to be a dynamical variable which varies in space and time, with scaling parameters to be found from the fit.  However, a more physical approach is to attempt to include the sources of the non-thermal pressure, be it sub-grid fluid motions, sub-grid magnetic fields, cosmic ray pressure, or some other phenomenon.  Including fluid motion in the baryon initialization as discussed in the previous point, may also reduce the need for non-thermal pressure to some extent.
\item While our results with inclusion of the magnetic field are encouraging, there is certainly room for improvement.  In particular, the coherence length of the initial magnetic field has an affect on the structure and growth of the magnetic fields during the collision, and this is being explored.  Furthermore, the required level of magnetic fields is related to the phenomena discussed above, which are also the sources of physical viscocity, so the reliability of the inferred magnetic fields can only be settled after these have been better understood.  
\end{enumerate}

Finally, more and better observational data will be a valuable addition to the modeling efforts. Our results depend heavily on having an accurate reconstruction of the current mass distribution.  Since the X-ray flux is proportional to the square of the baryon density, and the baryon density depends on the dark matter density, small changes in the mass distribution lead to large changes in the calculated X-ray flux.   Errors in the mass reconstruction will therefore lead to errors in our determination of the optimal initial conditions.  To constrain the magnetic fields, polarized radio emission and Faraday rotation measures of sources behind the bullet -- especially strongly lensed ones! --  would help the effort immensely.

\section{Conclusions}
\label{Conclusions}
We have presented a detailed simulation of the bullet cluster collision, which gives a 2D fit to observational data spanning an impressively wide range of wavelengths.  The major conclusions of this work are as follows:
\begin{itemize}
\item A simple initial configuration of two triaxial clusters fits the observed data for mass lensing quite well.  The triaxialities required are compatible with those seen in N-body simulations.  
\item The infall velocities required to explain the observations are not excessive compared to $\Lambda$CDM simulations.  Further comparison to cosmological predictions will be presented in a companion paper, in preparation.
\item The observed X-ray flux is quite sensitive to the baryon structure and can be reproduced with reasonable assumptions for the initial baryon distributions, although the fits are not as good as the fits to the mass lensing data only.  The best-fit structure of the initial clusters is similar to other observed clusters.  %The best-fit baryon fraction within the clusters is close to the average in the Universe, but slightly higher; whether this is an artifact of our simplified description is under investigation.   
\item A significant amount of non-thermal pressure, roughly equal to the thermal pressure, is required in order to fit plasma temperature and S-Z effect observations.
\item Addition of magnetic fields to the simulations improves the simultaneous fit of the mass lensing and X-ray flux data, and the magnetic fields required are in rough agreement with what is required to explain the radio halo observations.  %Work is ongoing to determine whether robust predictions are possible for the magnetic field and relativistic electrons.  
\end{itemize}

\section{Acknowledgements}
We have benefited from innumerable valuable discussions and advice from a great many colleauges -- too many to list them all.   Volker Springel and Yuval Birnboim provided support for getting started with the initial simulations.  Eugene Vasiliev provided the SMILE software, and we thank him for that and for helpful discussions on its use.  Much of the data analysis was done using the code known as yt \cite{yt2011}, which greatly simplifies analysis of AMR output.  Thanks also to Marusa Bradac for providing the mass lensing data, Tom Plagge for providing the SZ data and Maxim Markevitch for providing the extracted temperature map.  Most of the simulations were done under GID S1248 on the NASA Pleiades supercomputer system, for which we gratefully acknowledge support.  This work has been supported in part by grants NNX08AG70G, NSF PHY-1212538, NSF PHY-0900631 and NSF PHY-0970075.

\appendix
\section{Appendix - Simulation Details}
This appendix describes some of the details of the simulation tools and simulation conditions that are used.
\subsection{Simulation Conditions}
\label{Simulation_Conditions_Section}
We evaluated two simulation tools for this work, Gadget \cite{Gadget}, an SPH code, and Enzo \cite{Enzo}, a grid-based hydro code with adaptive mesh refinement (AMR).  Both simulators use discrete particles for the dark matter, but differ in the simulation of the hydrodynamics.  The two simulators were found to give similar results, but ultimately, the Enzo simulator was chosen when the need for incorporation of MHD into the simulations became apparent.  A summary of the key simulation conditions is shown in Table \ref{Simulation_Conditions}. 
\begin{table}[H]
	\centering
	\begin{tabular}{|c|c|c|} 
		\hline 
		Parameter & Value & Units \\ 
		\hline 
		\hline 
		Simulation Volume & 12000 x 6000 x 6000 & kpc \\ 
		\hline 
		Coarse Grid & 128 x 64 x 64 & - \\ 
		\hline 
		Maximum Number of Refinement Levels & 6 & - \\ 
		\hline 
		Minimum Grid Cell Size & 5.8 & kpc \\ 
		\hline 
		Total Number of Grid Cells & ~3.2E6 & - \\ 
		\hline 
		Maximum Baryon Mass per Grid Cell & 2.5E8 &  $M_\odot$ \\ 
		\hline
		Number of DM Particles & 5.0E6 & - \\ 
		\hline
		Mass of DM Particles & ~1.5E8 &  $M_\odot$ \\ 
		\hline
		Hydro Method & Runge Kutta 3rd-order MUSCL w/ Dedner MHD & - \\ 
		\hline
		Cosmology & Flat, Static & - \\ 
		\hline
	\end{tabular} 
	\caption{Summary Enzo simulation conditions.}
	\label{Simulation_Conditions}
\end{table}

\subsection{Resolution}
\label{Resolution}
\begin {figure}[H]
	\centering
	\subfigure[Resolution reduced by 4x]{\includegraphics[trim = 4.02in 0.0in 0.7in 0.34in, clip, width=0.31\textwidth]{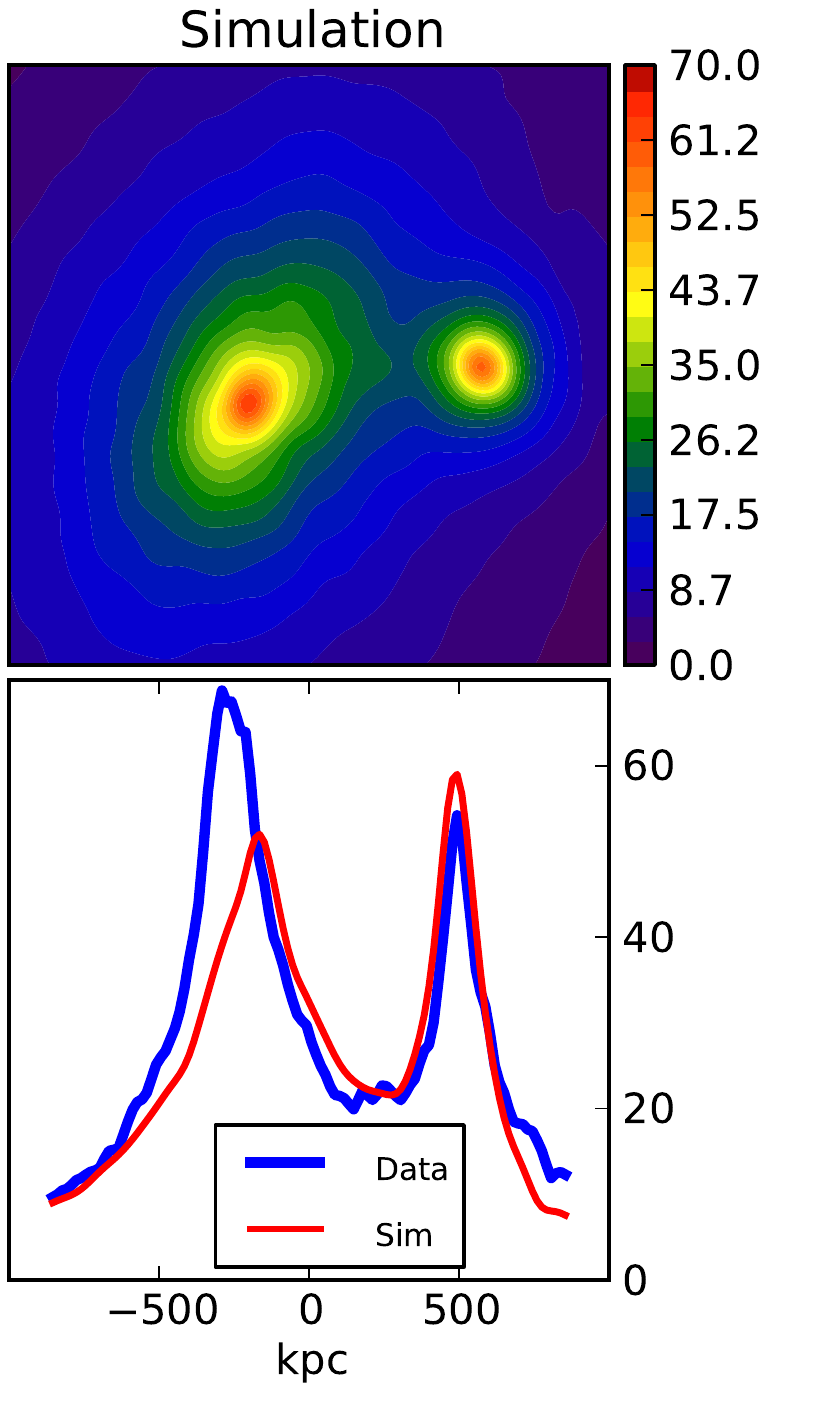}}
	\subfigure[Standard Resolution]{\includegraphics[trim = 4.02in 0.0in 0.7in 0.34in, clip, width=0.31\textwidth]{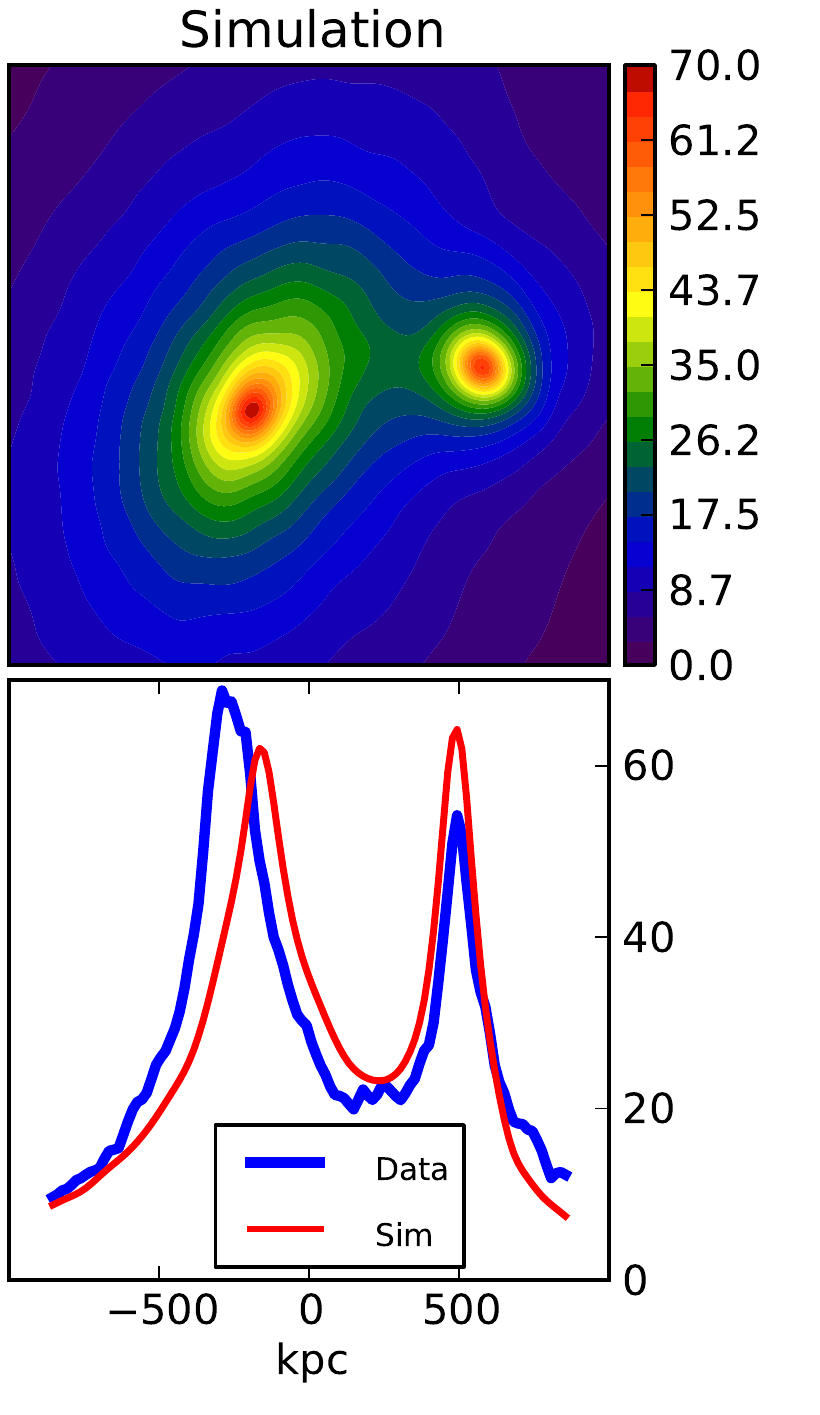}}
	\subfigure[Resolution increased by 2x]{\includegraphics[trim = 4.02in 0.0in 0.7in 0.34in, clip, width=0.31\textwidth]{Best_Fit01.pdf}}
  \caption {Impact of resolution on Mass Lensing fit}
  \label{Resolution_Impact_1}
  % Standard resolution is from multi_matrixn76/run2 Hi resolution is from batch_matrixn80/batch2/run28B Lo resolution is from multi_lowresn140/run1
  \end{figure}

\begin {figure}[H]
	\centering
	\subfigure[Resolution reduced by 4x]{\includegraphics[trim = 4.02in 0.0in 0.7in 0.34in, clip, width=0.31\textwidth]{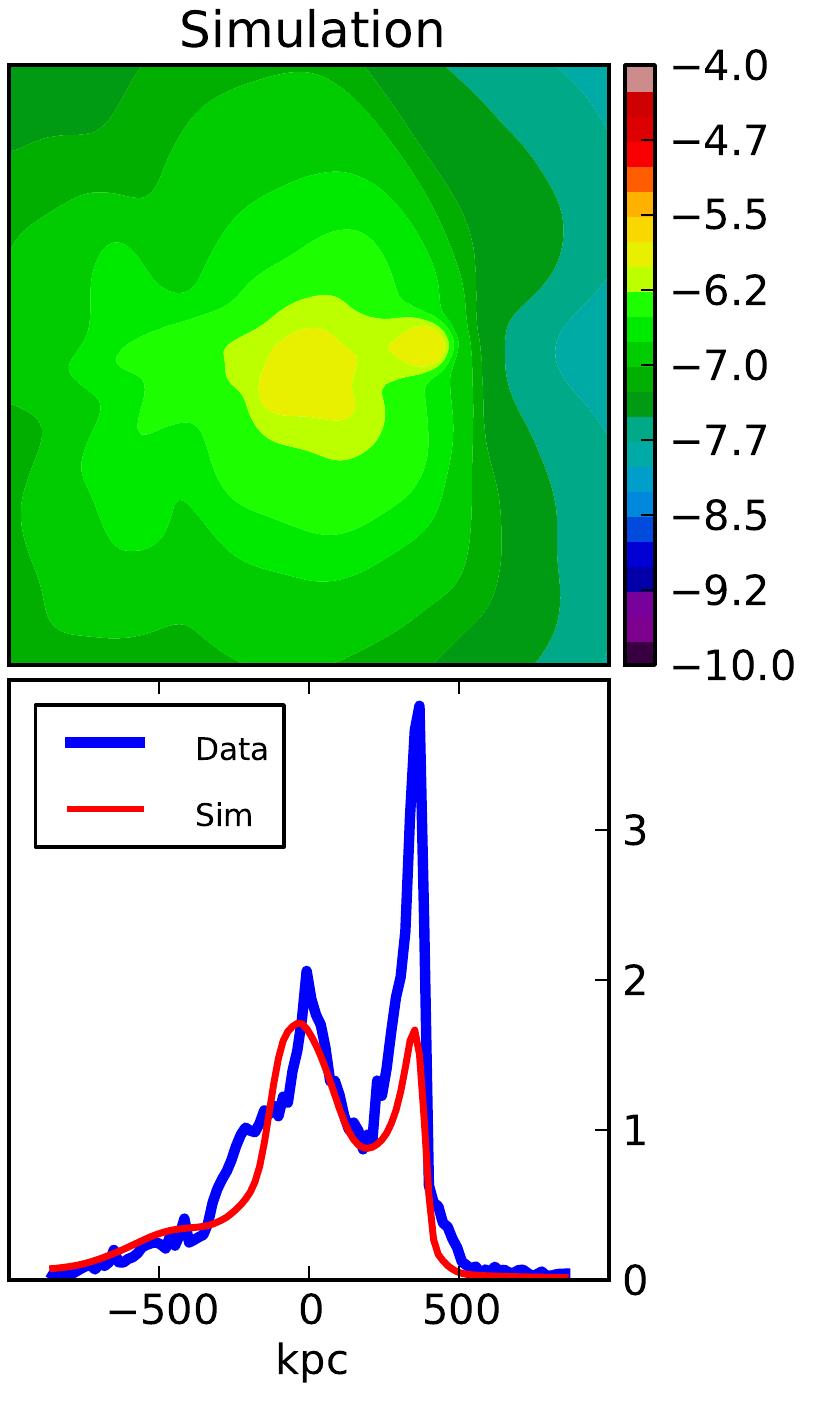}}
	\subfigure[Standard Resolution]{\includegraphics[trim = 4.02in 0.0in 0.7in 0.34in, clip, width=0.31\textwidth]{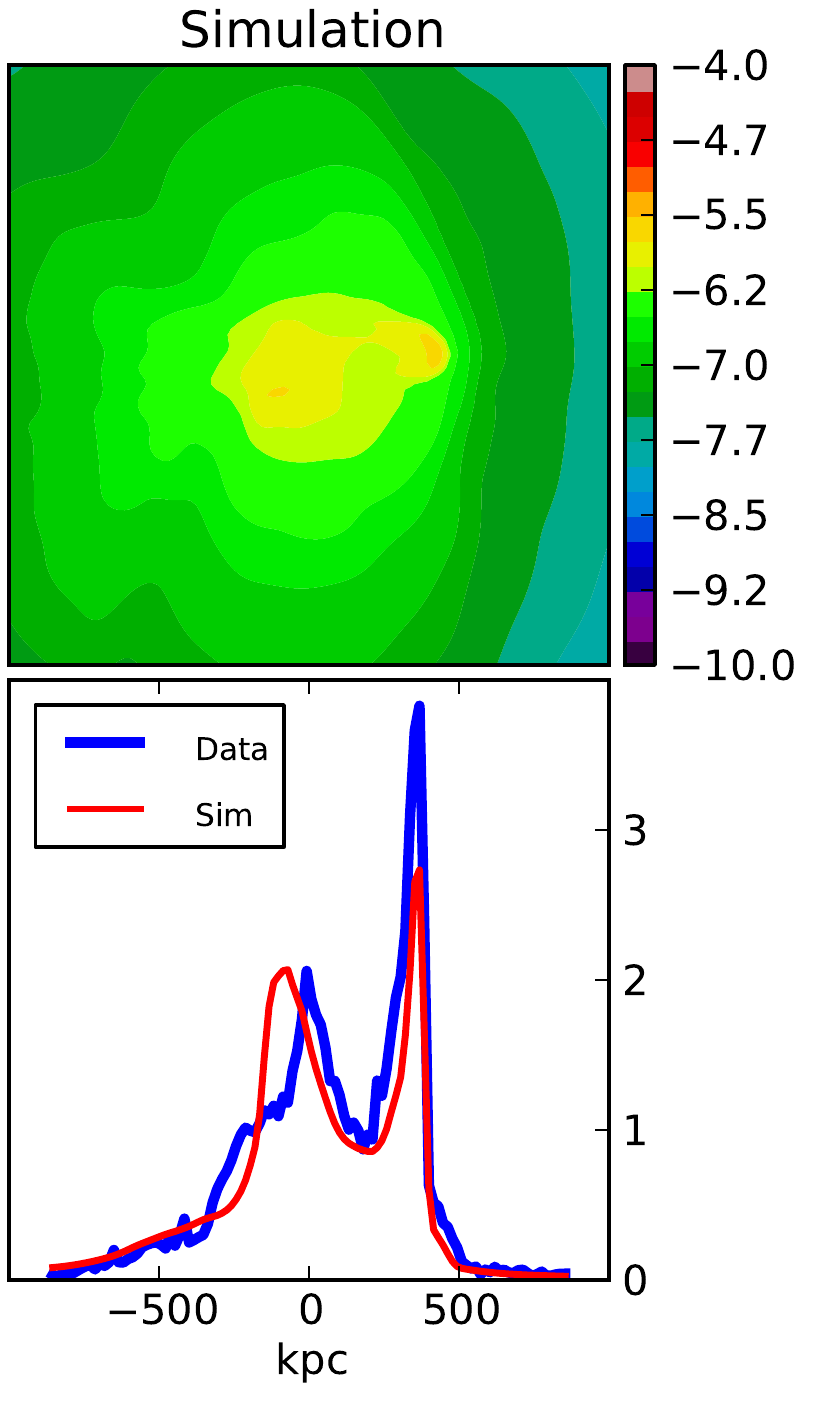}}
	\subfigure[Resolution increased by 2x]{\includegraphics[trim = 4.02in 0.0in 0.7in 0.34in, clip, width=0.31\textwidth]{Best_Fit02.pdf}}
  \caption {Impact of resolution on X-ray intensity fit}
  \label{Resolution_Impact_2}
  % Standard resolution is from multi_matrixn76/run2 Hi resolution is from batch_matrixn80/batch2/run28B Lo resolution is from multi_lowresn140/run1
  \end{figure}

To verify that the simulations are of sufficiently high resolution to capture the main features of the cluster collision, simulations using the optimal initial conditions are run at higher and lower resolutions.  For the lower resolution simulation, the number of dark matter particles is reduced by a factor of four, and the minimum grid size is increased by a factor of two.  For the higher resolution simulation, the number of dark matter particles is increased by a factor of two, and the minimum grid size is decreased by a factor of two.  All of these values are relative to the nominal values in Table \ref{Simulation_Conditions}.  Figures \ref{Resolution_Impact_1} and \ref{Resolution_Impact_2} summarize the results.  While there are some slight changes, especially in resolving the right-hand X-ray peak, the basic features of the simulation are unchanged, confirming that the main conclusions are not impacted by the resolution of the simulation.

\subsection{Calculation of Observables}
\label{Calculation_of_Observables}
This section describes the procedures used to calculate the observables from the simulation variables.  All cosmological calculations assume the following parameters:
	\begin{equation}
	\rm
	H_0 = 70 \frac{km}{sec\hspace{2mm}Mpc}\hspace{10mm}\Omega_m = 0.30\hspace{10mm}\Omega_\Lambda = 0.70 .
	\end{equation}
With these parameters, the Bullet Cluster redshift of z = 0.296 implies a luminosity distance of 1.53 Gpc.
\subsubsection{Lensing Mass} 
Calculation of the lensing mass for comparison to the lensing data is straightforward.  The baryon mass is a conserved quantity in the simulation, while the dark matter consists of discrete particles of fixed mass.  We sum the dark matter and baryon mass separately along the line of sight for each pixel, then add the two together to give the total lensing mass.  From the standpoint of this calculation, stars are indistinguishable from dark matter.  This should be a good approximation because the stellar mass accounts for only a few percent of the total baryonic mass in extremely massive clusters \cite{Gonzalez2013}.

\subsubsection{Calculation of X-ray Flux}
\label{Xray_Flux_Section}
\begin{figure}[H]
	\centering
	\includegraphics[trim = 0.7in 3.8in 1.0in 1.0in, clip, width=\textwidth]{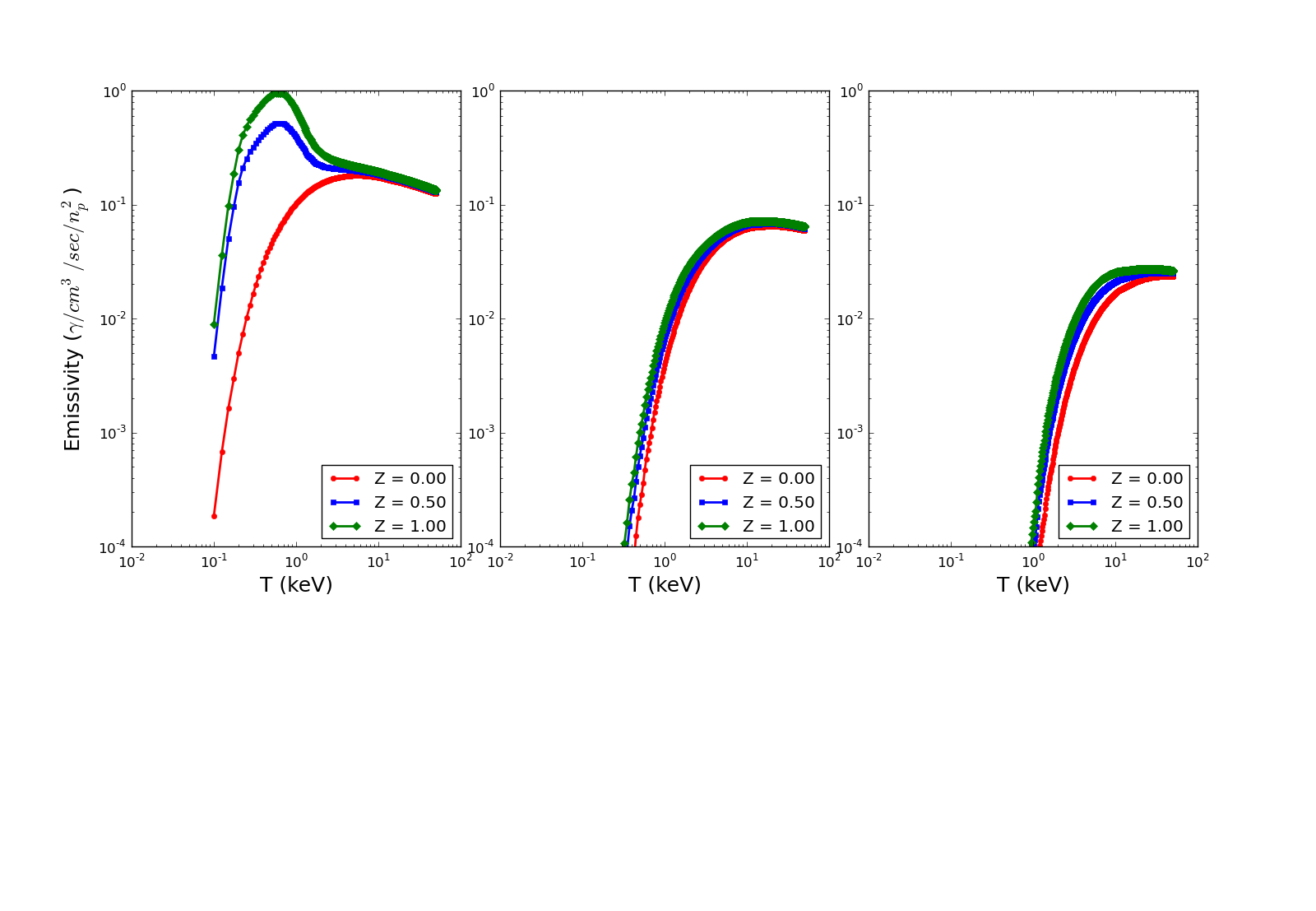}
	% Trim is Left Bottom Right Top
	\caption{Plasma emissivity in $\rm photons/cm^3/s/(n_p^2)$ where $\rm n_p$ is the plasma mass density, as calculated using the APEC code for different metallicity choices.  The metallicity Z is given as a fraction of solar. The left, center, and right plots are for energy bins of 500eV - 2000eV, 2000eV - 5000eV, and 5000eV - 8000eV, respectively. }
	\label{Xray_Flux}
	% This figure is from bullet/data/apec
	\end{figure}

As discussed above, the calibrated Chandra data is the flux of X-ray photons in $\rm photons/(cm^2 sec)$ in the given band of energies.  In the temperature range of these plasmas (approximately 1-50 keV), the radiation includes both thermal bremsstrahlung and line emission, and is a complex function of temperature and metallicity.  In this work, we use the APEC code \cite{ciao06} to build a look-up table of the plasma emissivity in each energy bin (with the energy bin limits blue-shifted back to the source appropriately) as a function of temperature and metallicity; the resulting emissivity is shown in Figure \ref{Xray_Flux}.  Since the plasma is optically thin, we calculate the total flux by summing the flux from each volume element along the line of sight.

We use the same calculation of plasma emissivity to calculate the rate of gas cooling during the simulation by building a look-up table that gives the cooling rate as a function of temperature and metallicity.  Because emission at all energies contribute to the plasma cooling, we sum the X-ray emissivity across all energies for this calculation.
\subsubsection{Plasma Temperature} 
Calculation of a 2D map of the plasma temperature to compare to observed temperatures is not trivial, since both the plasma density and the plasma temperature vary along the line of sight.  In principle the X-ray flux should be integrated along the line of sight, and the resulting spectrum fit with a temperature for each pixel, but this is computationally expensive.  We find that a weighted average temperature using a weighting function equal to the X-ray flux  gives almost the same result, and is much faster, so we use this procedure to produce the temperature maps (for example, Figure \ref{Best_Fit_Temp}).  In other words, we calculate the temperature in each pixel by
	\begin{equation}
	\rm
	T = \frac{\int_{\zeta_{min}}^{\zeta_{max}} T(\zeta)*\epsilon(n_p(\zeta),T(\zeta),Z)d\zeta}{\int_{\zeta_{min}}^{\zeta_{max}} \epsilon(n_p(\zeta),T(\zeta),Z)d\zeta} , 
	\end{equation}
where the integration is along the line of sight, $\rm \zeta_{min}, \zeta_{max}$ are the boundaries of the simulation volume, and the plasma emissivity $\rm \epsilon(n_p,T,Z)$ is calculated as described above.  We use $\rm E_{min}$ and $\rm E_{max}$ values of 500 eV and 8000 eV for this purpose. 

\subsubsection{Calculation of S-Z Effect}
Inverse Compton scattering of CMB photons by the hot plasma leads to a distortion of the CMB blackbody spectrum.  A good approximation for optically thin, non-relativistic plasmas such as these is that the distortion results in a slight modification of the CMB temperature as given by the following equation \cite{Birkinshaw}:
	\begin{equation}
	\rm
	\frac{\Delta T}{T_{CMB}} = -2\sigma_T \int \frac{k_B T(\zeta)}{m_e c^2}n_e(\zeta) d\zeta .
	\end{equation}
Here the integration is along the line of sight, and $\sigma_T$ is the Thomson scattering cross section.  Since the X-ray flux is proportional to $\rm n_e^2$ and is relatively independent of T, and the SZE $\rm \Delta T$ is proportional to $\rm n_e T$, the two datasets together allow an independent determination of plasma density and temperature.  For the Bullet Cluster, the S-Z effect has a maximum $\rm \Delta T$ of approximately $\rm 400 \mu K$.

\subsubsection{Radio Halo} 
\label{Radio_Calculation}
Galactic clusters, especially those undergoing violent collisions, are known to have extended radio halos.  The source of the radio emission is less well understood than the source of the X-ray emission, but is believed to be a population of relativistic electrons which emit synchrotron radiation as they spiral around the cluster magnetic field lines \cite{Govoni}.  Following closely Rybicki and Lightman \cite{R-L}, we use the following model:
\begin{itemize}
	\item The population of relativistic electrons follows a power law distribution (we assume $ p>2$):
	\begin{equation}
	{\rm
	N(\gamma) d\gamma = C} \gamma^{-p} d\gamma .
	\label{ESpectrum}
	\end{equation}
	\item The population of relativistic electrons is in equipartition with the magnetic field, meaning that:
	\begin{equation}
	{\rm
	\int_0^\infty \gamma m_e c^2 C} \gamma^{-p} d\gamma = \frac{\rm B^2}{8\pi} .
	\end{equation}
\end{itemize}
This implies:
	\begin{equation}
	{\rm
	C =} \frac{(p-2) \rm B^2}{8\pi \rm m_e c^2} .
	\end{equation}
	
With these assumptions, the intensity of radio emission depends only on the magnetic field intensity B and the power law exponent $p$, which we assume constant throughout the simulation volume. After averaging over the randomly oriented direction of the magnetic field, the radio power is given by:
	\begin{equation}
	\rm
	P(\nu)d\nu = \frac{\sqrt{3 \pi}}{32\pi^2 (1+z) }\frac{e^3 B^3}{(m_e c^2)^2}\hspace{4 pt}f({\it p})\hspace{4 pt}(\frac{2\pi m_e c \nu(1+z)}{3 e B})^{-({\it p}-1)/2} d\nu ,
	\end{equation}
	where the function f({\it p}) is given by the following expression:
	\begin{equation}
	{\rm
	f}(p) = \frac{(p-2)}{(p+1)}\frac{\Gamma(\frac{p+5}{4})}{\Gamma(\frac{p+7}{4})}\Gamma(\frac{p}{4}+\frac{19}{12})\Gamma(\frac{p}{4}-\frac{1}{12}) ;
	\end{equation}
 f$(p)\approx 0.3$ for typical values of  $ p ~(2.5<p<4.0)$.  For comparing to observations, it is important to note that the spectral index of the electron energy, $p$, (Equation \ref{ESpectrum}) is related to the spectral index of radio emission, $s$,
	\begin{equation}
	{\rm
	f(\nu) d\nu = C} \nu^{-s} d\nu ,
        \label{SIndex}
	\end{equation}
by the following equation \cite{R-L}:
	\begin{equation}
	%\rm
	s = \frac{p-1}{2} .
        \label{P_s_Equation}
	\end{equation}

We find the power law exponent to be tightly constrained to a value of $p \approx 3.6$, as discussed in Section \ref{Mag_Field_2}.

\clearpage
\bibliographystyle{unsrt}
\bibliography{bullet}
\end{document}